\documentclass[pdftex,twocolumn,epjc3]{svjour3}          

\RequirePackage[T1]{fontenc}

\smartqed  

\RequirePackage{graphicx}
\RequirePackage{mathptmx}      
\RequirePackage{flushend}
\RequirePackage[numbers,sort&compress]{natbib}
\RequirePackage[colorlinks,citecolor=blue,urlcolor=blue,linkcolor=blue]{hyperref}

\journalname{Eur. Phys. J. C}

\begin{document}

\title{Extensive beam test study of prototype MRPCs for the T0 detector at the CSR external-target experiment\thanksref{t1}}


\author{Dongdong Hu\thanksref{addr1,addr2}
        \and
        Jiaming Lu\thanksref{addr1,addr2} 
        \and
        Jian Zhou\thanksref{addr1,addr2}
        \and
        Peipei Deng\thanksref{addr1,addr2}
        \and
        Ming Shao\thanksref{e1,addr1,addr2}
        \and
        Yongjie Sun\thanksref{e2,addr1,addr2}
        \and
        Lei Zhao\thanksref{e3,addr1,addr2}
        \and
        Hongfang Chen\thanksref{addr1,addr2}
        \and
        Cheng Li\thanksref{addr1,addr2}
        \and
        Zebo Tang\thanksref{addr1,addr2}
        \and
        Yifei Zhang\thanksref{addr1,addr2} 
        \and
        Yi Zhou\thanksref{addr1,addr2}
        \and
        Wenhao You\thanksref{addr1,addr2}
        \and
        Guofeng Song\thanksref{addr1,addr2}
        \and
        Yitao Wu\thanksref{addr1,addr2}
}

\thankstext[$\star$]{t1}{Thanks to the title}
\thankstext{e1}{e-mail: swing@ustc.edu.cn}
\thankstext{e2}{e-mail: sunday@ustc.edu.cn}
\thankstext{e3}{e-mail: zlei@ustc.edu.cn}

\institute{State Key Laboratory of Particle Detection and Electronics, University of Science and Technology of China, Hefei 230026, China\label{addr1}
          \and
         Department of Modern Physics, University of Science and Technology of China, Hefei 230026, China\label{addr2}
          \and
          \emph{Present Address:}No.96, JinZhai Road Baohe District,Hefei,Anhui, 230026,P.R.China\label{addr3}
}

\date{Received: date / Accepted: date}

\maketitle

\begin{abstract}
The CSR External-target Experiment (CEE) will be the first large-scale nuclear physics experiment device at the Cooling Storage Ring (CSR) of the Heavy-Ion Research Facility in Lanzhou (HIRFL) in China. A new T0 detector has been proposed to measure the multiplicity, angular distribution and timing information of charged particles produced in heavy-ion collisions at the target region. Multi-gap resistive plate chamber (MRPC) technology was chosen as part of the construction of the T0 detector, which provides precision event collision times (T0) and collision geometry information. The prototype was tested with hadron and heavy-ion beams to study its performance. By comparing the experimental results with a Monte Carlo simulation, the time resolution of the MRPCs are found to be $\sim$ 50 ps or better. The timing performance of the T0 detector, including both detector and readout electronics,  we found to fulfil the requirements of the CEE. 
\end{abstract}

\section{Introduction}

One of the main purposes of heavy-ion collisions is to study the bulk properties of strongly interacting matter and understand the quantum chromo dynamics (QCD) phase diagram \cite{myers1998nuclear}. At finite temperature (T) and chemical potential ($\mu$), QCD describes relevant features of nuclear physics in the early universe, in neutron stars and in heavy ion collisions. By varying the collision energy, different nuclear matter states and phase structure can be exploited. In the region of low temperature and high net baryon density, the nuclear equation of state (EOS) is most important for understand the phase diagram, gaining a better understanding of the properties of stellar objects and heavy nuclei \cite{braun2007quest,zhang2010searching,McLerran:2003yx}, and confirming the possible occurrence of a hypothetical quarkoyonic matter phase at very high baryon number density \cite{mclerran2009quarkyonic,andronic2010hadron,feng2010probing,xie2013symmetry,sumiyoshi1994relativistic,steiner2005isospin}. More theoretical and experimental efforts are definitely required to arrive at a convincing constraint of E$_{sym}$($\rho$) \cite{xiao2014probing}. Dedicated heavy-ion experiment at energies of several hundred AMeV will help resolve these large theoretical uncertainty, which is well covered by the Cooling Storage Ring (CSR) of the Heavy-Ion Research Facility in Lanzhou (HIRFL).\par
The CSR external-target experiment (CEE) has been proposed to study (1) the density dependence of nuclear symmetry energy by measuring the $\pi^-$/$\pi^+$ ratio (and other relevant observables) for various heavy-ion collision systems, (2) the EOS at supra-saturation density, and (3) the rich QCD phase at high-density and low-temperature. The CSR\cite{xiao2009nuclear} can deliver a wide range of heavy-ion beams from deuteron (up to 1 AGeV) to uranium nuclei (up to 520 AMeV), and therefore,  can provide significant opportunity to study E$_{sym}$($\rho$) and the properties of cold nuclear matter and quarkoyonic matter phase at very high baryon number density. For example, the $\pi^-$/$\pi^+$ ratio in heavy-ion collision in this energy region can be a sensitive probe\cite{xiao2014probing1}.\par
\begin{figure}[htbp]
\begin{minipage}[t]{1\linewidth}
\centering
\includegraphics*[width=60mm]{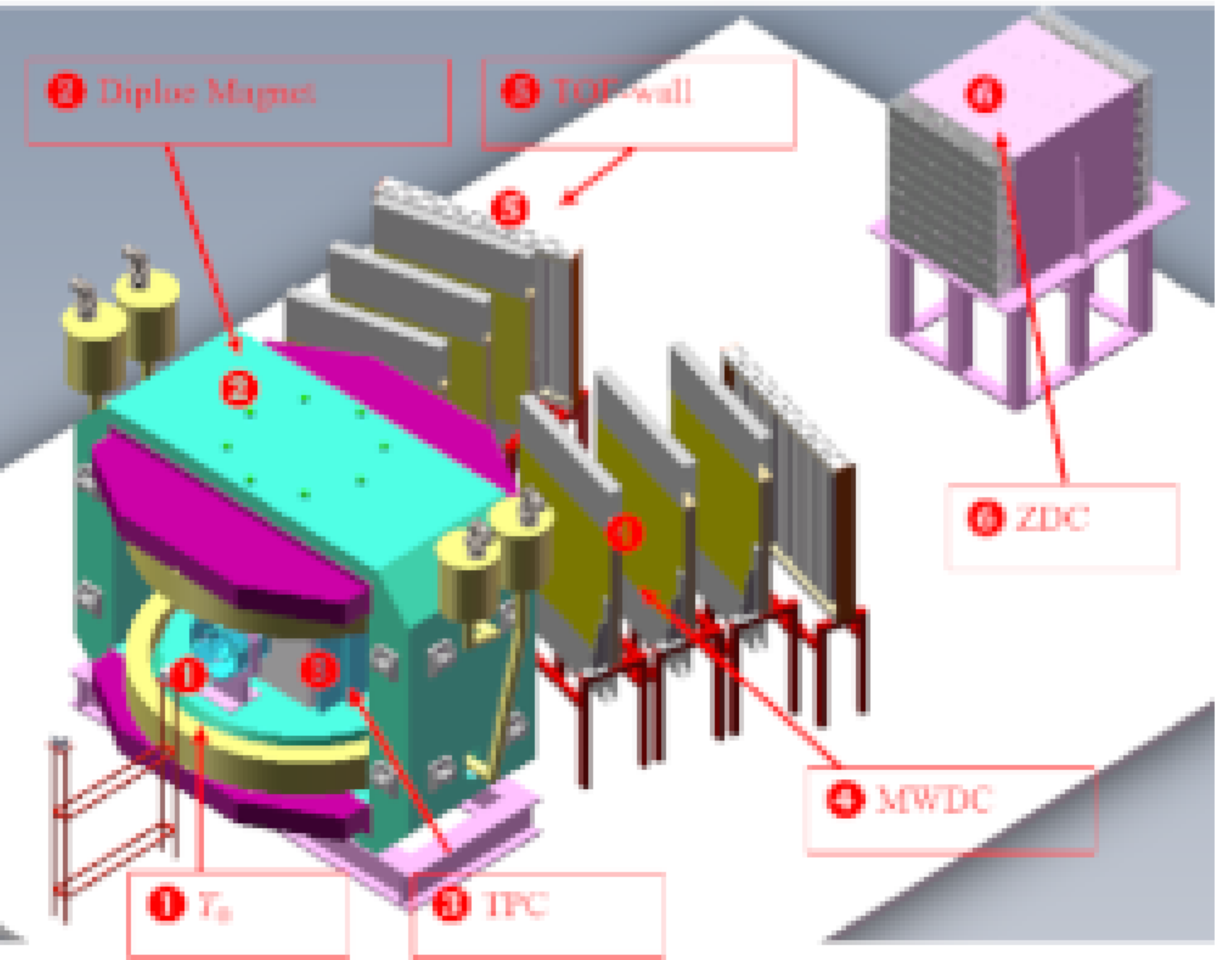}
\caption{Schematic layout of the CEE detector system, taken from ref. [15].}
\label{cee}
\end{minipage}
\vfill
\begin{minipage}[t]{1\linewidth}
\centering
\includegraphics*[width=60mm]{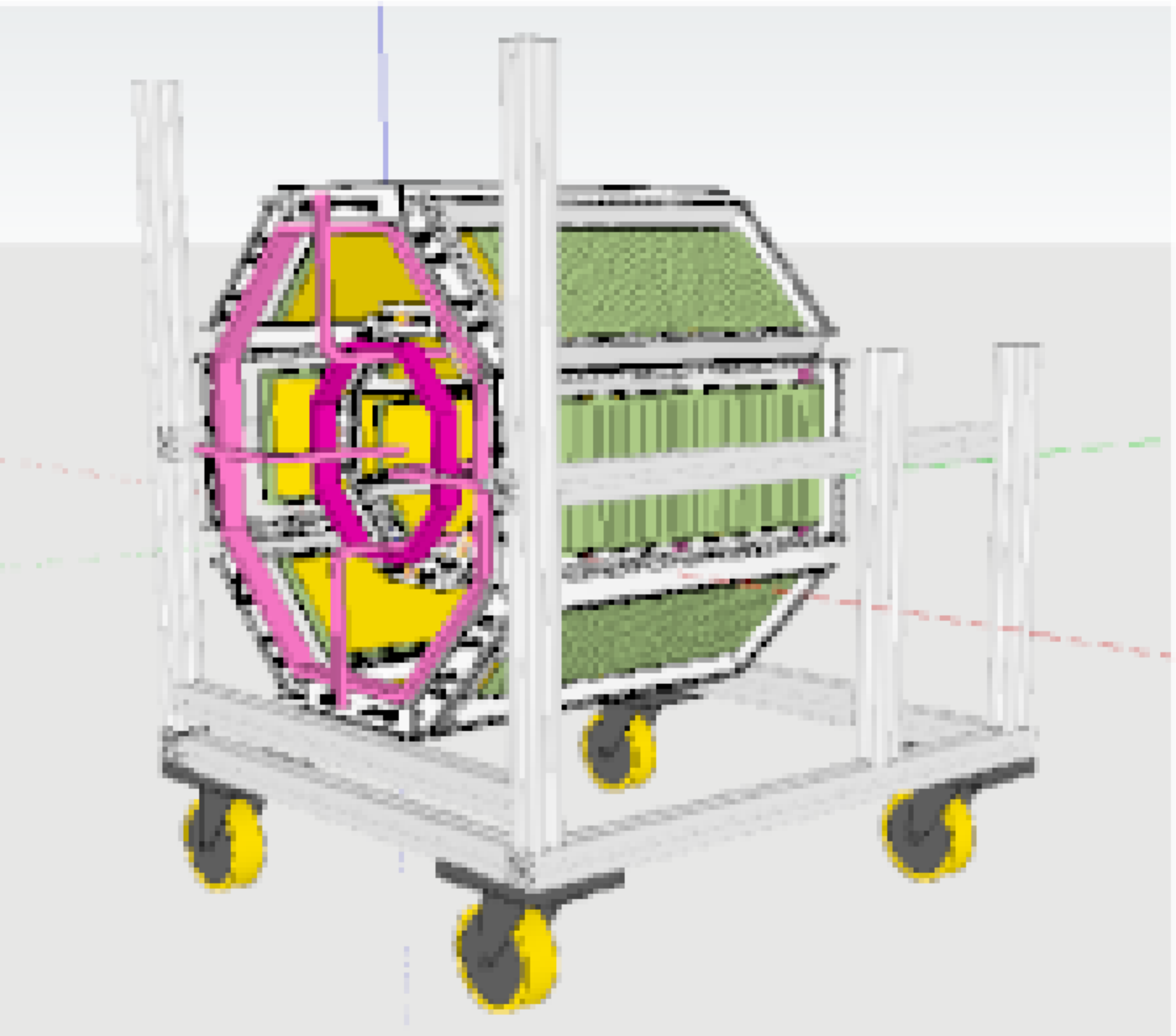}
\caption{Sketch of the structure of T0 detector, which is composes of inner and outer layers. Each layer includes eight MRPCs.}
\label{magnet}
\end{minipage}
\end{figure}

The CEE system includes a large angle dipole magnet, tracking detectors, a Time of Flight (TOF) system, and a zero-degree calorimeter (ZDC), as depicted in Fig. \ref {cee} \cite{Lu2016}. The TOF system contains a T0 detector, an internal TOF (iTOF) and an external TOF (eTOF). The T0 detector is located $\sim$10 centimeter distance around the target region to detect the final-state charged particles and clusters in the heavy-ion reaction. Both the target and the T0 detector sit in a strong magnetic field. The TOF technique is employed to identify the charged final-state particles. As the start detector of the TOF system, the T0 detector not only determines the collision time with high precision but also serves as a trigger detector for the experimental system, providing information on the event multiplicity and reaction plane. \par
\begin{figure}[htb]
\centering
\includegraphics*[width=80mm]{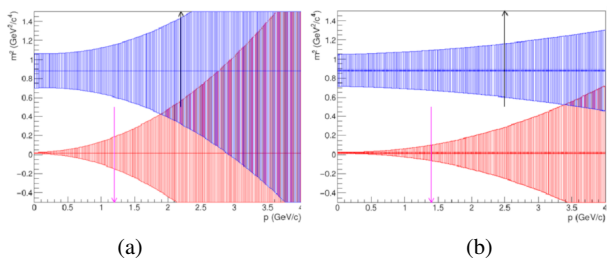}
\centerline{(a)    \qquad   \qquad  \qquad   \qquad   \qquad    \qquad    \qquad                                     (b)}
\caption{$m^2$ distribution vs. particle momentum, measured by the CEE iTOF (a) and eTOF (b). In both plots, the red and blue areas identify bands within 2$\sigma$ of the $m^2$ distribution of pions and protons as a function of momenta. The pink arrow marks the upper momentum under which 99.5\% of the pions reside. The black arrow has similar meaning, but for protons.}
\label{1a}
\end{figure}

In the GEANT4 simulation, a kinetic energy 1.0AGeV Ar-Ar collisions were generated by UrQMD3.4.6\cite{A:2005}, and $\sim$10,000 UrQMD data generated events were fed to the MC detector system to test the TOF performance. The time difference between the collision point and the hit on TOF detector was recorded. TOF timing uncertainties of 50, 100, 150 and 200 ps were studied. A 5\% smearing to the particle momentum was added to account for the reconstruction uncertainty. Track length uncertainties of 0.5 and 2cm were also included in the simulation for particles hitting the iTOF and eTOF, respectively. The simulation results are shown in Fig.\ref {1a}, for both the iTOF and eTOF, with the TOF time resolution setting at 100ps. In the plots, the red and blue areas denote bands within 2$\sigma$ of the m$^2$ distribution of pions and protons as a function of momenta. The pink arrow marks the upper momentum under which 99.5\% of the pions reside. The black arrow has a similar meaning for protons. There were very few kaon in the final state, so they were neglected in the plots. It is clear from the figure that the pion/proton separation is easier for eTOF because of the much longer flight path (>2.5m) than iTOF (0.5$\sim$1.2m), and a TOF system with an overall time resolution of 100ps is quite adequate for pion/proton identification for the CEE. \par
It’s noted that in Fig.\ref {1a} the 100 ps time resolution includes intrinsic contribution from both T0 detector and iTOF and eTOF detectors. For example, a 70 ps time resolution for both the T0 and iTOF detector combined to roughly 100 ps.  
In our design the intrinsic time resolution for the T0 detector needed to be < 80 ps.
 leading to 80 ps for the T0 detector.

\section{The module design}
Fig.\ref {magnet} shows the schematic structure of the T0 detector, which consists of eight inner and eight outer MRPCs\cite{gapienko2013studying,yong2012prototype,zeballos1996new,spegel2000recent}. In this design, a time resolution <80 ps and an efficiency >95\% are sought for the CEE-T0. Later in this paper, we describe the efforts to build and test a T0 detector prototype, the details of the configuration of MRPCs, the readout electronics, the hadron and heavy ion beam test, and the results obtained from the beam test.
\subsection{The configuration of MRPC}
\begin{figure}[htbp]
\centering
\begin{minipage}[t]{1.0\linewidth}
\centering
\includegraphics*[width=80mm]{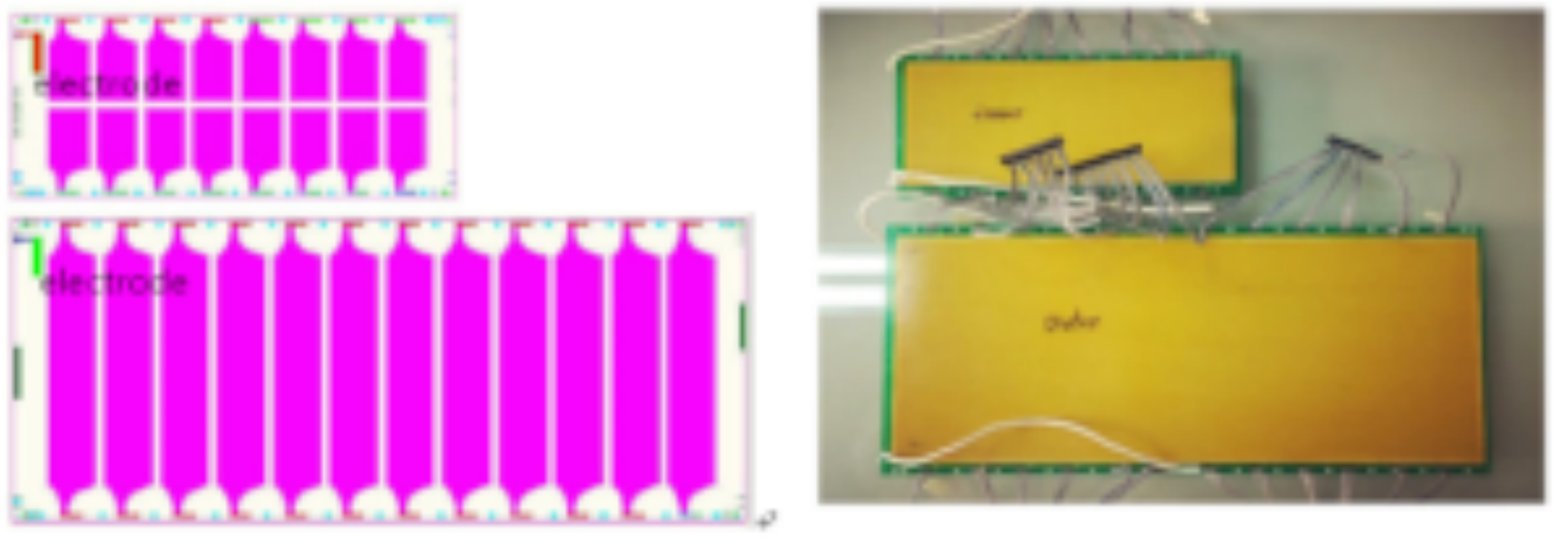}
\centerline{(a)    \qquad   \qquad    \qquad  \qquad    \qquad (b)}
\end{minipage}
\vfill
\begin{minipage}[t]{1.0\linewidth}
\centering
\includegraphics*[width=80mm]{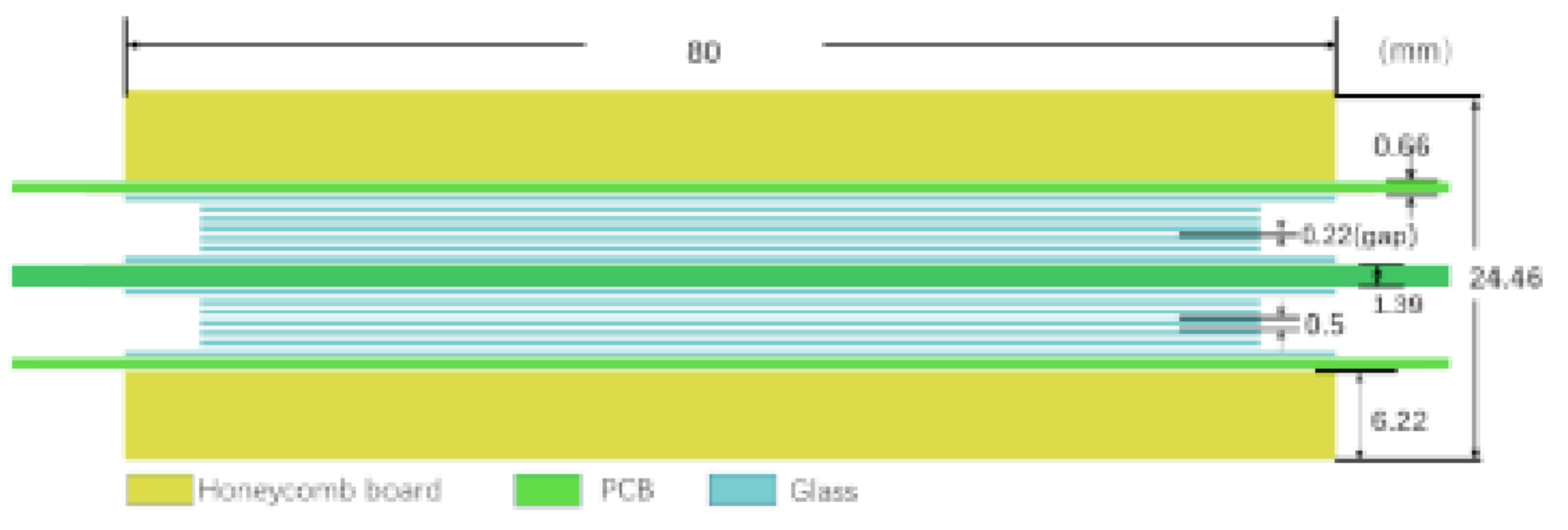}
\centerline{(c)}
\end{minipage}
\caption{MRPC module design and prototypes. The readout pad design for the inner layer is shown in the left image(a) at the top left, and the design for outer layer of the MRPC is shown in the lower image (a) at the top left. A photograph of the finished MRPC prototypes is shown at the top right (b). The side view of the MRPC is shown in the bottom panel (c).}
\label{mrpc}
\end{figure}

The inner and an outer layers of the T0 detector are composed of eight MRPCs, which are suitable for high precision timing and fast triggering. Fig. \ref {mrpc} shows design of the two kinds of MRPCs, for inner and outer layers respectively. The smaller inner MRPCs shown at the top of Fig.\ref {mrpc} (a) contain 16 single-end differential readout pads, each pad  3.05 cm long and 2.15 cm wide with a 0.35 cm gap, and the strips of larger outer MRPCs are each 12.0 cm long, 2.6 cm wide and segmented by 0.4 cm gap, with a total of 12 dual-end differential readouts. The sensitive volume of detector consists of  0.5 mm thick float glass plates consisting of a double-stack structure that is mirrored with respect to the central electrode (Fig. \ref {mrpc} bottom (c) ) with twelve gas gaps. High voltages (HV) is applied to the external electrodes surfaces. Each gap is separated by a nylon fishing line, with a diameter of 0.22 mm. Fig. \ref {mrpc} (upper right (b)) shows a photograph of the finished inner and outer MRPC modules.

\begin{figure}[htbp]
\centering
\begin{minipage}[t]{1.0\linewidth}
\centering
\includegraphics*[width=80mm]{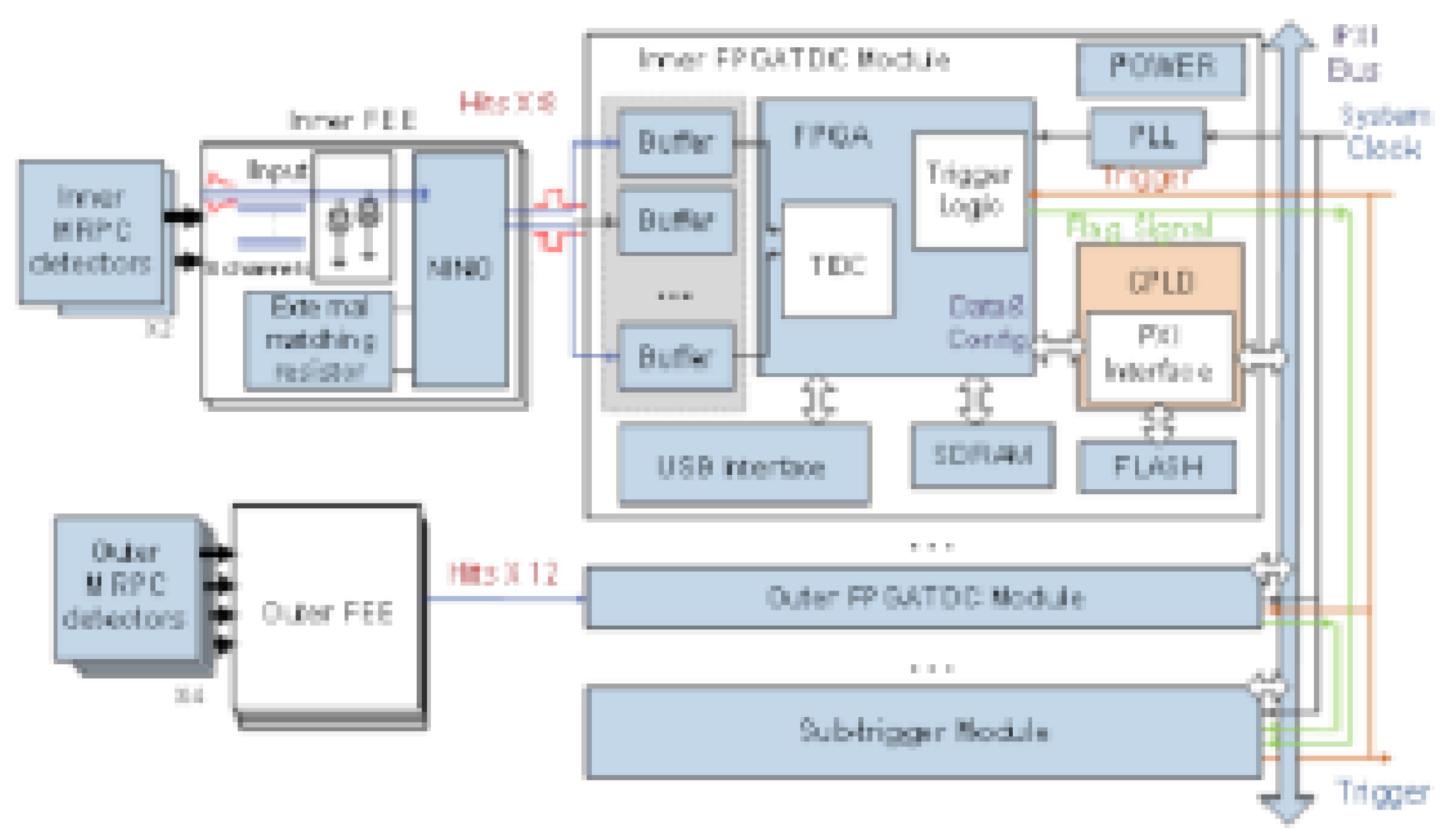}
\centerline{(a)}
\end{minipage}
\vfill
\begin{minipage}[t]{1.0\linewidth}
\centering
\includegraphics*[width=60mm]{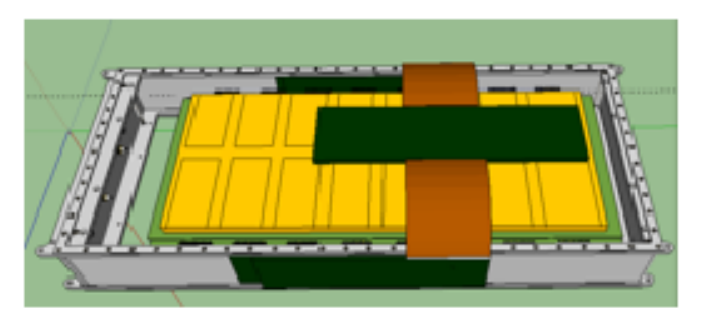}
\centerline{(b)}
\end{minipage}
\caption{Block diagram of the readout electronics and CSR T0 Module.}
\label{elect}
\end{figure}
\subsection{Readout electronics}
\begin{figure}[htb]
\centering
\includegraphics*[width=80mm]{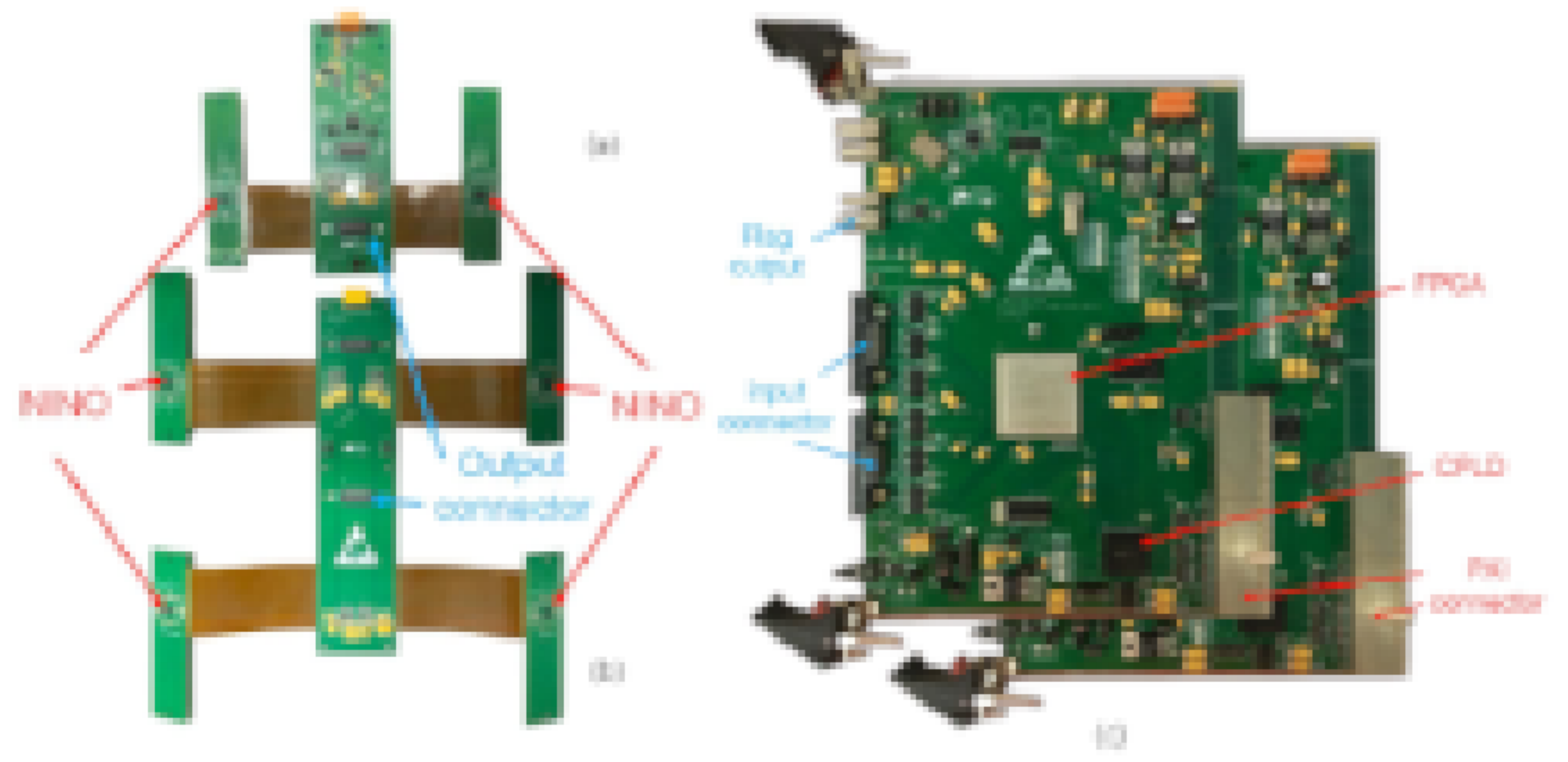}
\caption{NINO-based FEE modules for the inner (top a) and outer (bottom b) MRPC module. FPGA-based time-to-digital convertor is shown on the right. This FPGA TDC can achieve both leading and trailing edge time measurement with high precision. Trigger matching and other functions are also integrated.}
\label{elec1}
\end{figure}

The front-end electronics (FEEs) are located outside the gas box containing the MRPC module(Fig.\ref{elect} bottom (b)), which make use of the NINO chip\cite{anghinolfi2003nino}. This ultra-fast and low-power front-end amplifier/discriminator application specific integrated circuit(ASIC) was specially designed for the MRPC by the ALICE-TOF group. Fig. \ref {elec1} shows that the inner MRPC module's two NINO chips are used for 16 readout channels, and the outer MRPC module's four NINO chips are used to handle 24 channels. The off-chip resistor in the FEE (the “External matching resistor” in Fig. \ref {elect} top (a)) is used for impedance matching\cite{anghinolfi2004nino}, and each FEE module outputs corresponding LVDS signals with fast leading edges for timing purposes and the signal charge information contained in its width. The signal from FEE is then processed by the field programmable gate array-based time-to-digital convertor (TDC) module \cite{wang201110}. The feild programmable gate array (FPGA) TDC can achieve both leading and trailing edge time measurement in a single channel based on the carry chain structure within the FPGA slice resource with a time jitter of FPGA TDC <25 ps RMS for the leading edge. Trigger pre-processing, trigger matching based on CAM and DPRAM, and other functions are also integrated in one single FPGA device. The TDC module is designed based on PCI extensions for Instrumentation (PXI)-6U standard. The hardware configuration, data transfer, and online reconfiguration of the FPGA logic can be conducted by using a single board computer (SBC) located in Slot 0 through a PXI bus. A USB interface is also employed for system debugging. The block diagram of the readout electronics is shown at the top of Fig \ref {elect}. Fig. \ref {elec1} shows photographs of the FEE and FPGA TDC modules. After the readout electronics being designed and tested \cite{deng2018readout}, preliminary commissioning tests with the four T0-MRPC prototypes, including the two inner and two outer modules, were conducted in the laboratory with cosmic rays. Next, the whole system was fully tested with hadron and heavy-ion beams to study its performance in detail.

\section{Hadron in-beam test}

\subsection{IHEP-E3 beam line at BEPC II facility}
The IHEP-E3 beam line at the Beijing Electron-Positron Collider II (BEPCII), Institute of High-Energy Physics (IHEP), Beijing, China has been used to study the characteristics of the CEE-T0 MRPCs and the functionality of the new electronics. Secondary particles (mainly protons, $\pi^{-/+}$ and $e^{-/+}$ ) from an incident electron beam hitting a carbon target \cite{lijiacai2004bepc} were filtered and delivered to the IHEP-E3 line. Among these secondary particles, protons and pions were dominant. The particle momenta were tuned to 700 MeV/c. A Cherenkov detector (C0) and two scintillators (SC1 and SC2 with an overlapping area of 5 cm x 5 cm ) assembled by the XP2020 photomultiplier tubes were used for basic triggering. A coincidence of the two scintillators and an anti-coincidence with the Cherenkov detector allowed the selection of protons and pions \cite{besiii2009construction}. Three multi-wire chambers (MWPCs) were installed for the measurement of the beam trajectory. However, in this work, the Cherenkov detector and MWPC detectors were not included in the beam test.
\subsection{Test setup}
\begin{figure}[htbp]
\centering
\begin{minipage}[t]{1.0\linewidth}
\centering
\includegraphics*[width=70mm]{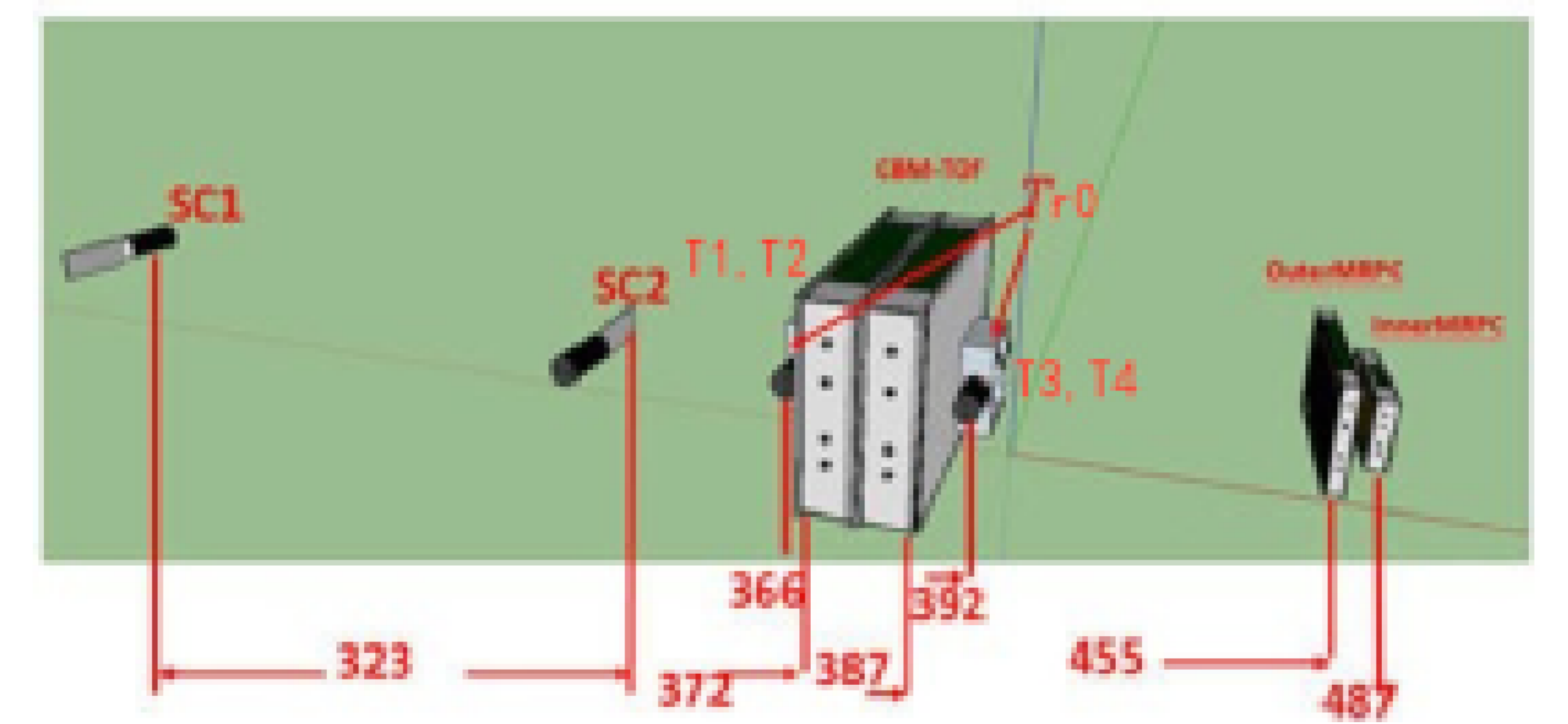}
\centerline{(a)}
\end{minipage}
\vfill
\begin{minipage}[t]{1.0\linewidth}
\centering
\includegraphics*[width=70mm]{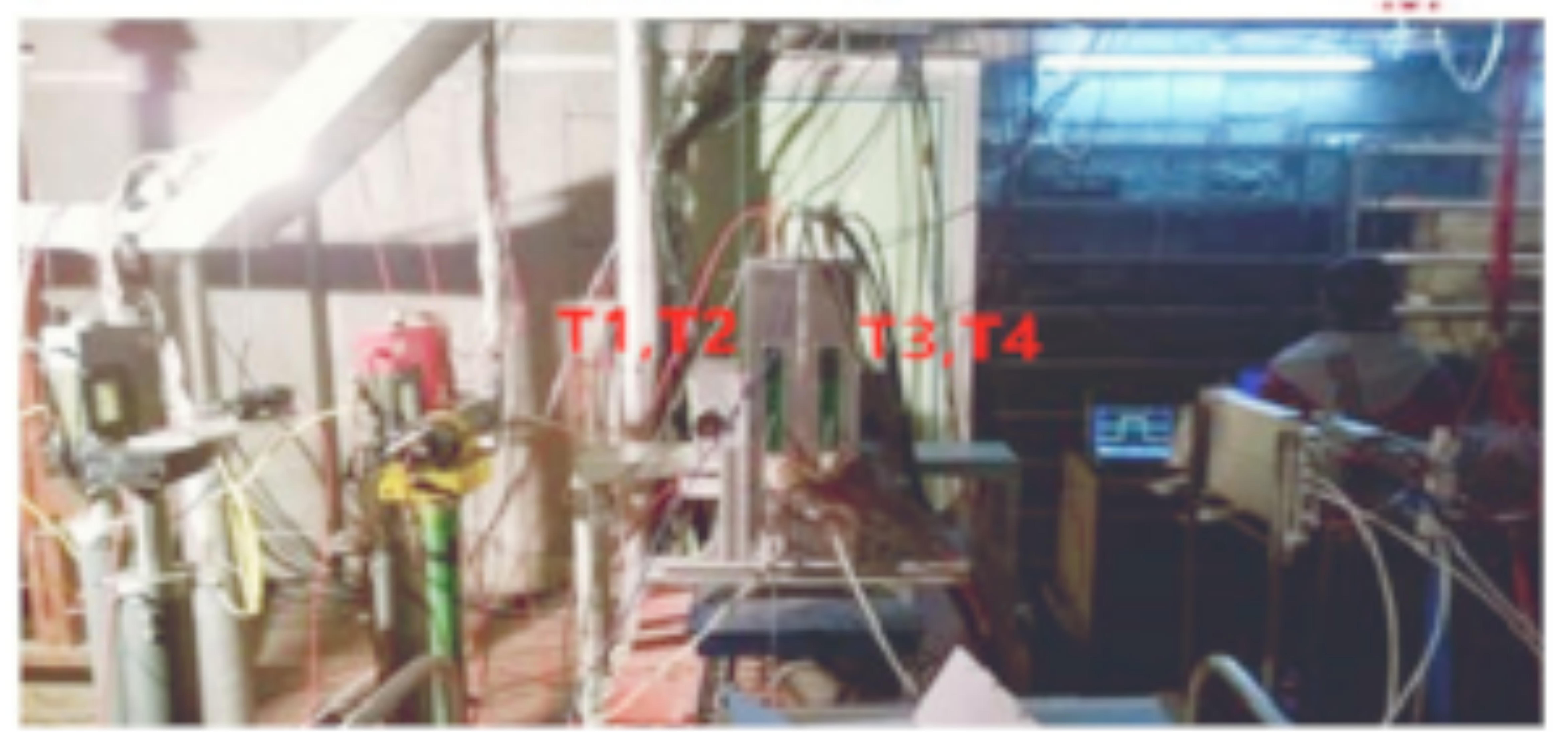}
\centerline{(b)}
\end{minipage}
\caption{Sketch (top a) and photograph (bottom b) of experimental setup at the E3 line of BEPCII. The numbers in the upper plot mark the position (in cm) along the beam line.}
\label{setup}
\end{figure}

A sketch of the in-beam test setup is shown at the top of Fig. \ref {setup}. The SC1 and SC2 scintillators provide the coincidence trigger, and the four small single-end readout scintillators (2 x 5 cm$^2$, BC420, divided into two groups: T1/T2 and T3/T4) coupled with fast photomultiplier tubes (PMT H6533) provide the accurate event reference time (Tr0=$\frac{(T1+T2+T3+T4)}{4}$). The inner and outer MRPC modules for CSR-T0 detector are placed with several other MRPC prototypes (CBM-TOF detectors), at the downstream position of the test beam. T1 and T2 were placed at the upstream position, and T3 and T4 were at the downstream position relative to the CBM-TOF MRPC modules. If a particle though SC1, SC2, T1, T2, T3, T4, its signals  are fed to a splitter. One copy is sent to HPTDC for precise timing measurement \cite{anghinolfi2003nino,christiansen2004hptdc}, while another output copy is fed to the discriminator. The HPTDC module, which has discrimination and signal transfer ability, was built by the USTC Electronics Group. After discrimination, the coincident signals of the SC1 and SC2 act as the trigger for the system. The digital signal from T1 to T4 act as the reference time start signals of the test system. The MRPC signals are amplified and discriminated by the FEE and then recorded by the FPGA-TDC. The difference between leading- and trailing-edge timing gives the signal width (time-over-threshold, (TOT)) information. The MRPC module were placed in a gas-tight aluminum box and flushed with a working gas mixture of 90\% R134a, 5\% iso-butane, and 5\% sulfur hexafluoride (SF$_6$). A photograph of the experimental setup is shown at the bottom of Fig. \ref {setup}. The operational parameters values were set according to the cosmic ray test results\cite{shao2008upgrade,hu2017t0}. In October 2016, two inner and two outer MRPCs for the CEE-T0 detector were tested at the IHEP-E3 line. The HV was set at $\pm$7200V, and the threshold was set to 220mV (i.e. 36 fC at the input of the NINO ASIC) for the test. The results and the analysis of the collected data are described in the following. 

\begin{figure}[htbp]
\begin{minipage}[t]{1.0\linewidth}
\centering
\includegraphics*[width=52mm]{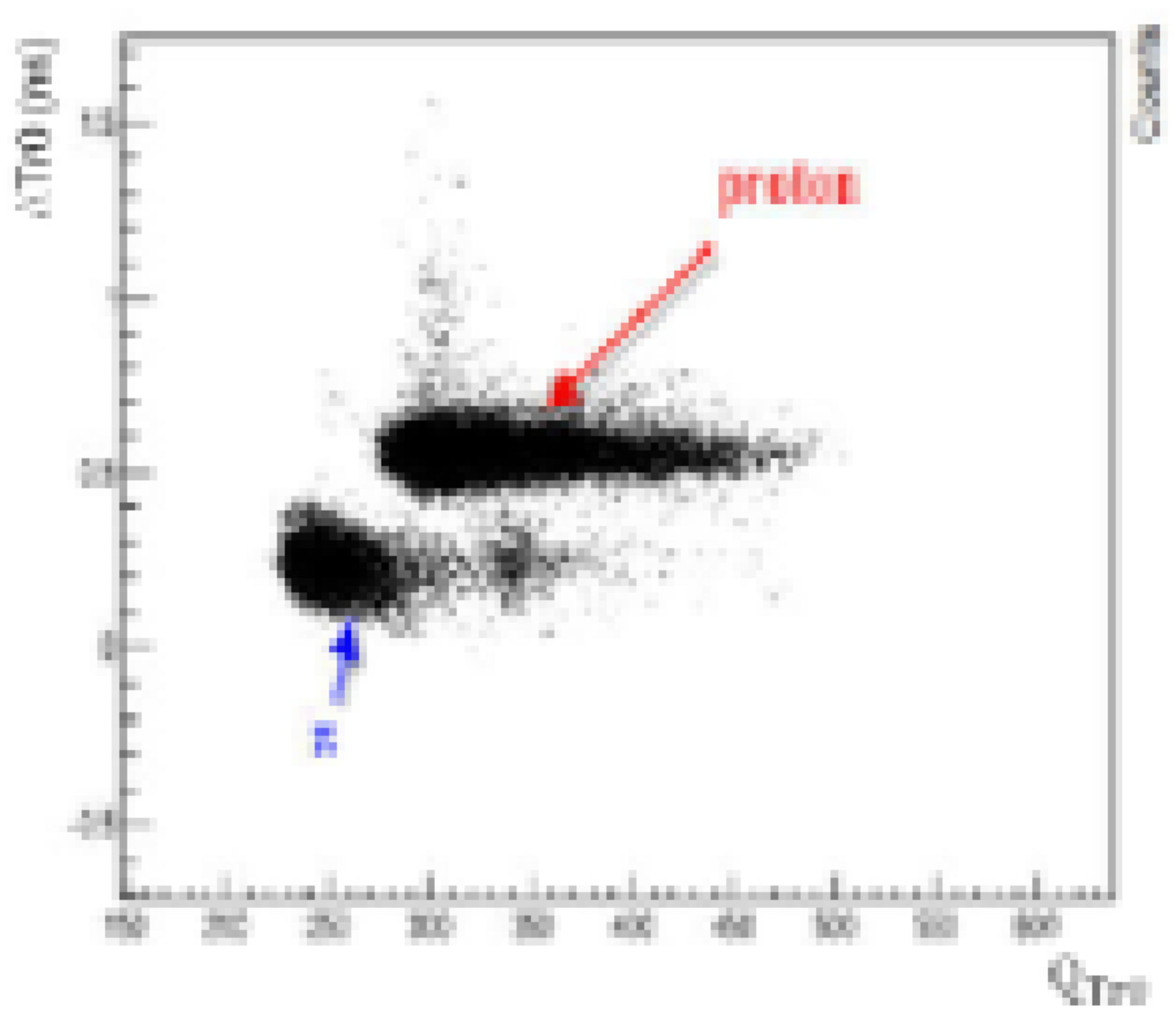}
\centerline{(a)}
\end{minipage}
\vfill
\begin{minipage}[t]{1.0\linewidth}
\centering
\includegraphics*[width=50mm]{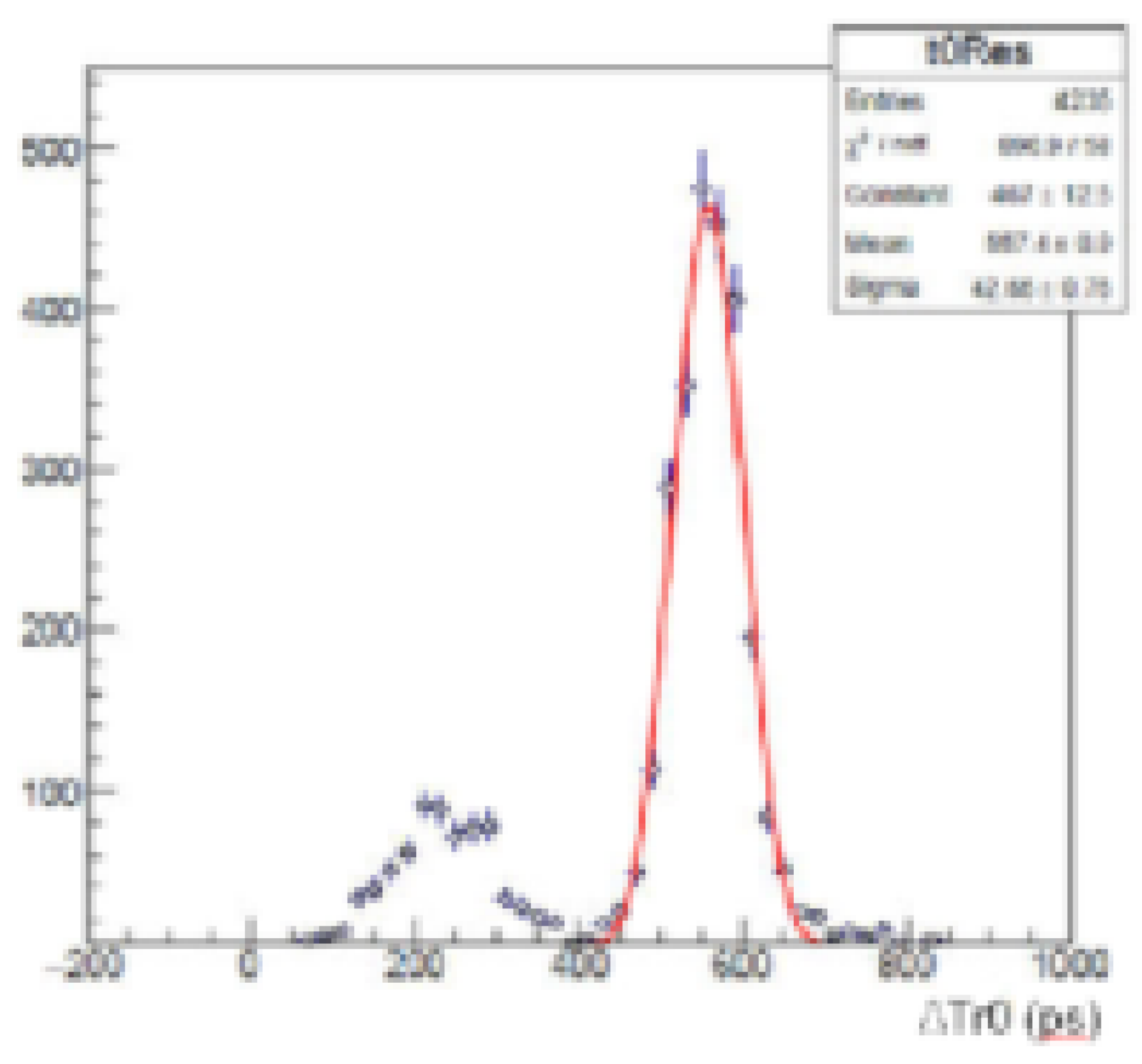}
\centerline{(b)}
\end{minipage}
\caption{(a) The $\Delta$Tr0 (=$\frac{(T1+T2-T3-T4)}{4}$) vs. signal charge information for pions and protons, and (b) the Tr0 distribution.}
\label{distiguish}
\end{figure} 
\subsection{Result of the in-beam test}
\subsubsection{Time resolution}
The IHEP-E3 beam was generated by bombarding a target with a primary electron beam such as Cu, Be or C. The secondary particles mix with e,$\pi$ and p, primarily protons and pions, so the first step of the analysis was to distinguish different particles. Fig. \ref {distiguish} (a) shows the scatter plot of time difference $\Delta Tr0$=$\frac{(T1+T2-T3-T4)}{4}$ between T1/T2 and T3/T4 and the signal charge of the Tr0 detector. At a beam momentum of 700MeV/c, protons deposit more energy in the Tr0 detector (scintillator) because of the larger dE/dx, and they travel more slowly than pions, resulting in larger $\Delta Tr0$ values across a defined path length (the distance between the T1/T2 pair and the T3/T4 pair is $\sim$26 cm, corresponding to $\Delta Tr0=\frac{L}{2cp}(\sqrt{p^2+m^2_{proton}c^2}-\sqrt{p^2+m^2_{\pi}c^2})$=282.6 ps). These influences are reflected in Fig.\ref {distiguish}. The separation of pions and protons can be clearly seen. Fig. \ref {distiguish} (b) shows the $\Delta Tr0$ distribution separately. By a Gaussian fit, we estimated the Tr0 time resolution to be $\sim$ 40 ps for proton beam. For pions, the Tr0 time resolution was a little larger, being close to 60 ps. \par
\begin{figure}[htbp]
\begin{minipage}[t]{1.0\linewidth}
\centering
\includegraphics*[width=50mm]{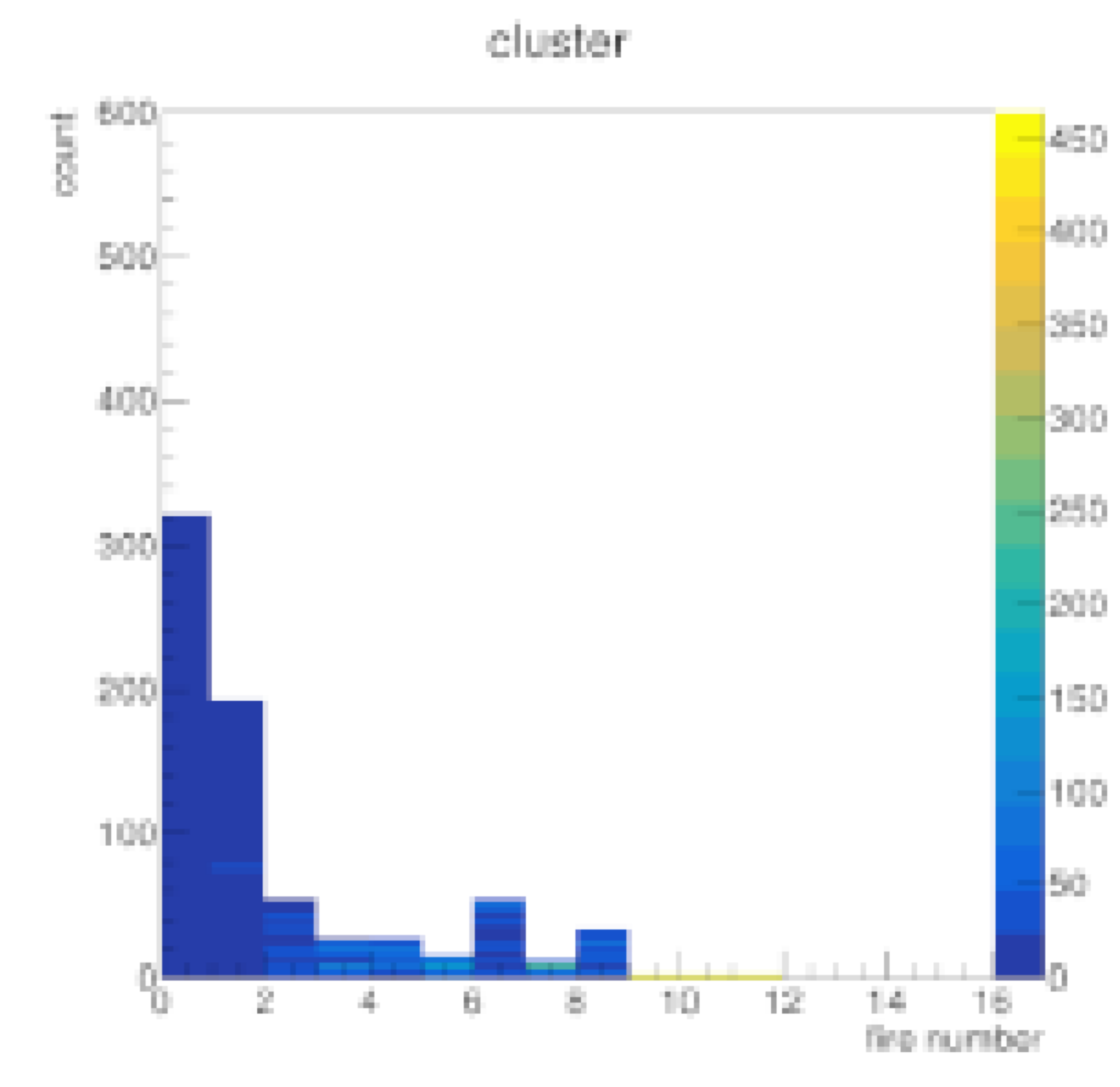}
\caption{Clustersize of the MRPC in hardon in-beam test.}
\label{cluster}
\end{minipage}
\vfill
\begin{minipage}[t]{1.0\linewidth}
\centering
\includegraphics*[width=50mm]{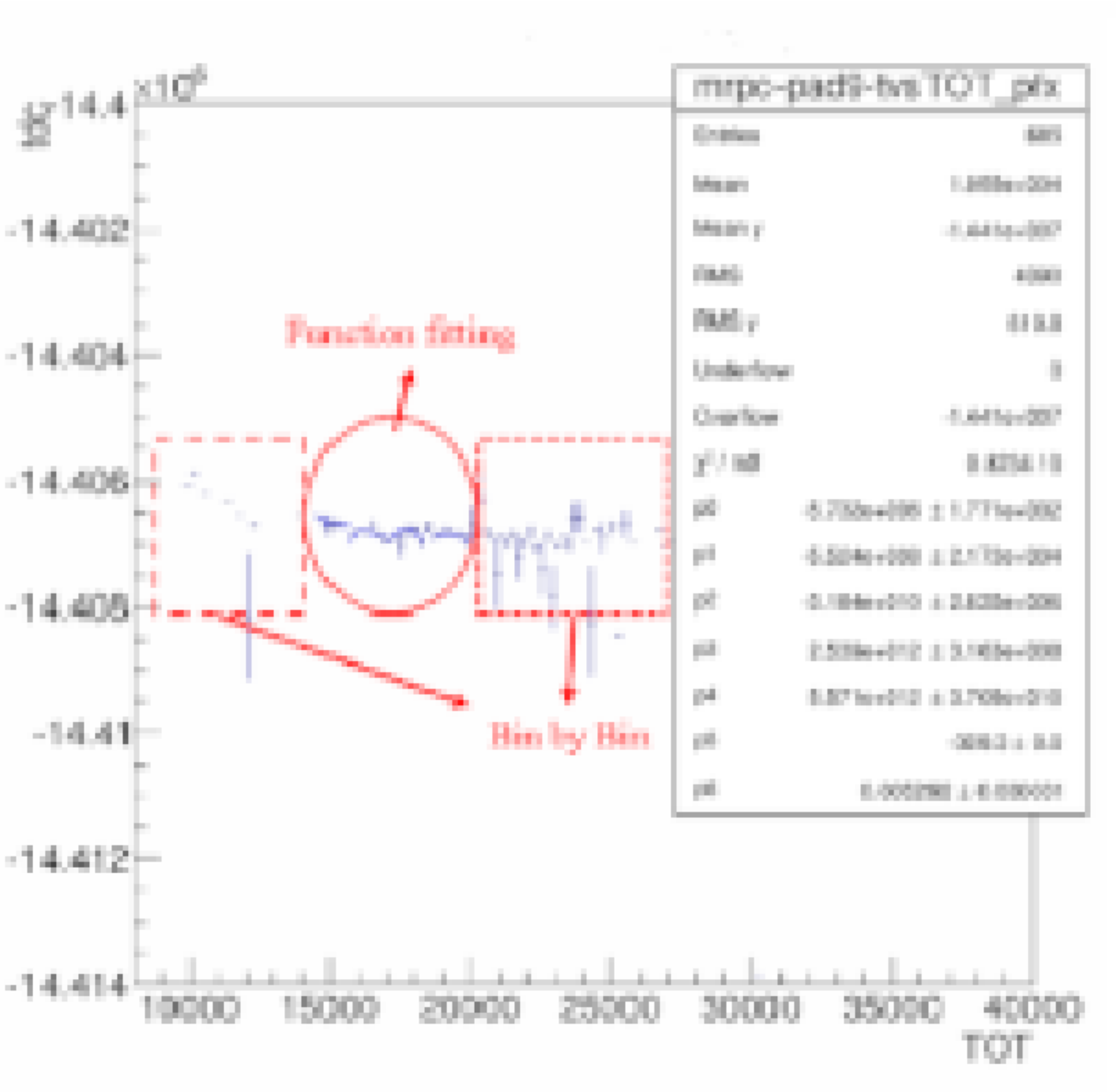}
\caption{MRPC timing and TOT plot divided into several parts.The circled part uses the function fitting and the rectangle parts use a bin-by-bin method to do slewing.}
\label{cluster1}
\end{minipage}
\end{figure}
\begin{figure}[htb]
\centering
\includegraphics*[width=50mm]{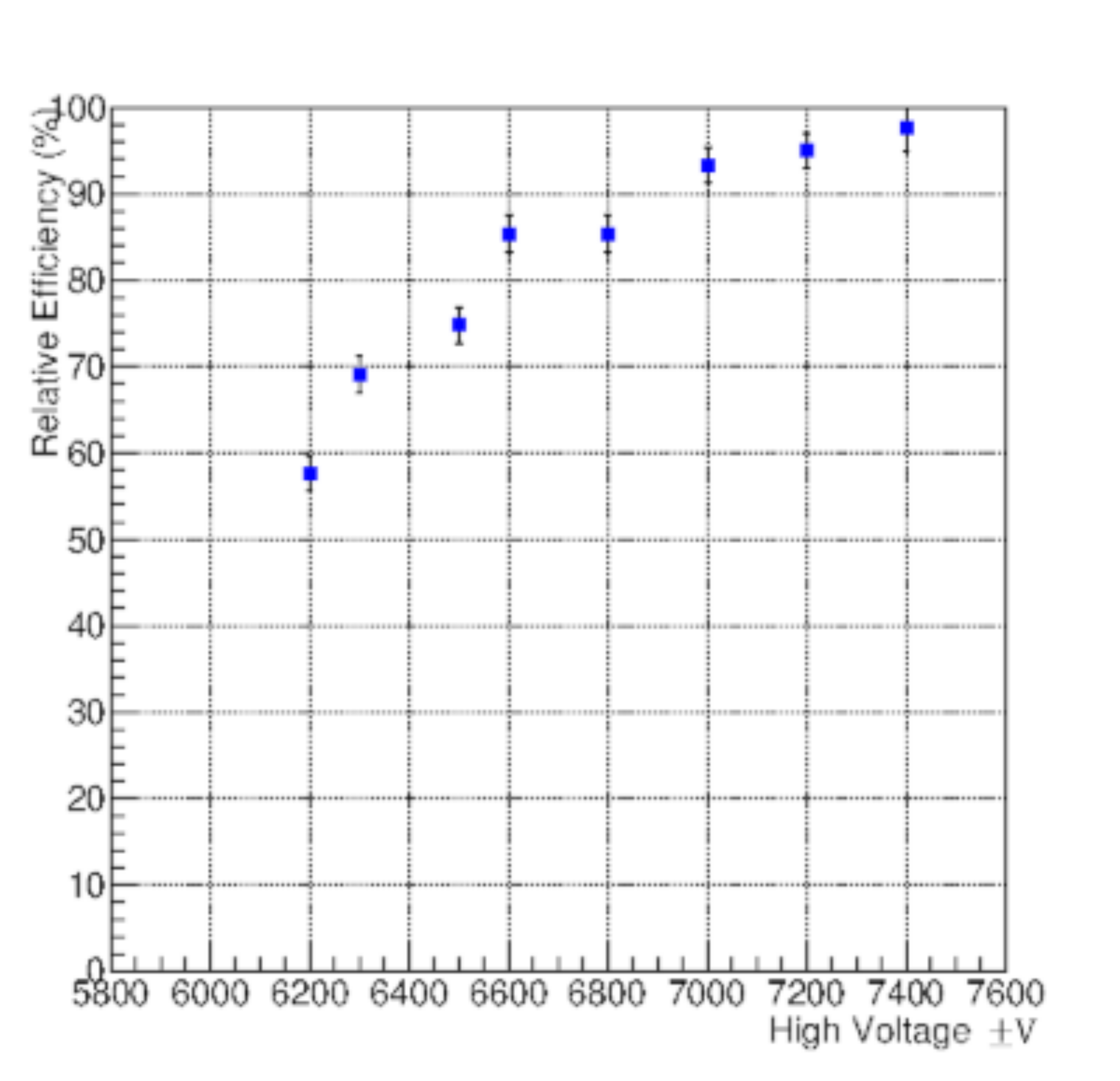}
\caption{Efficiency plateau at BEPCII.}
\label{eff}
\end{figure}
\begin{figure}[htb]
\centering
\includegraphics*[width=60mm]{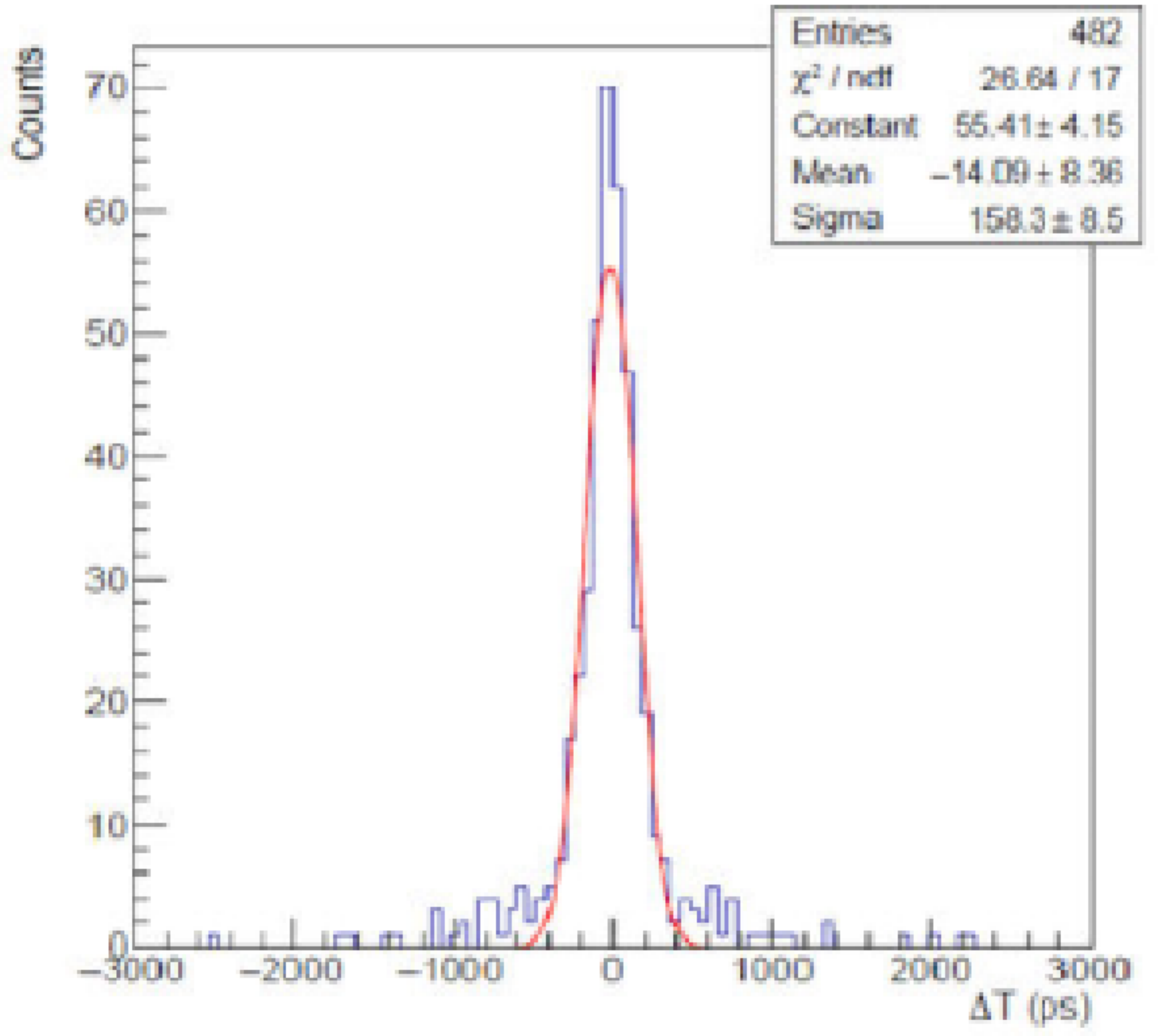}
\caption{MRPC time distribution relative to T$_{0}$ for a proton beam.}
\label{time}
\end{figure}
\begin{figure}[htb]
\centering
\includegraphics*[width=60mm]{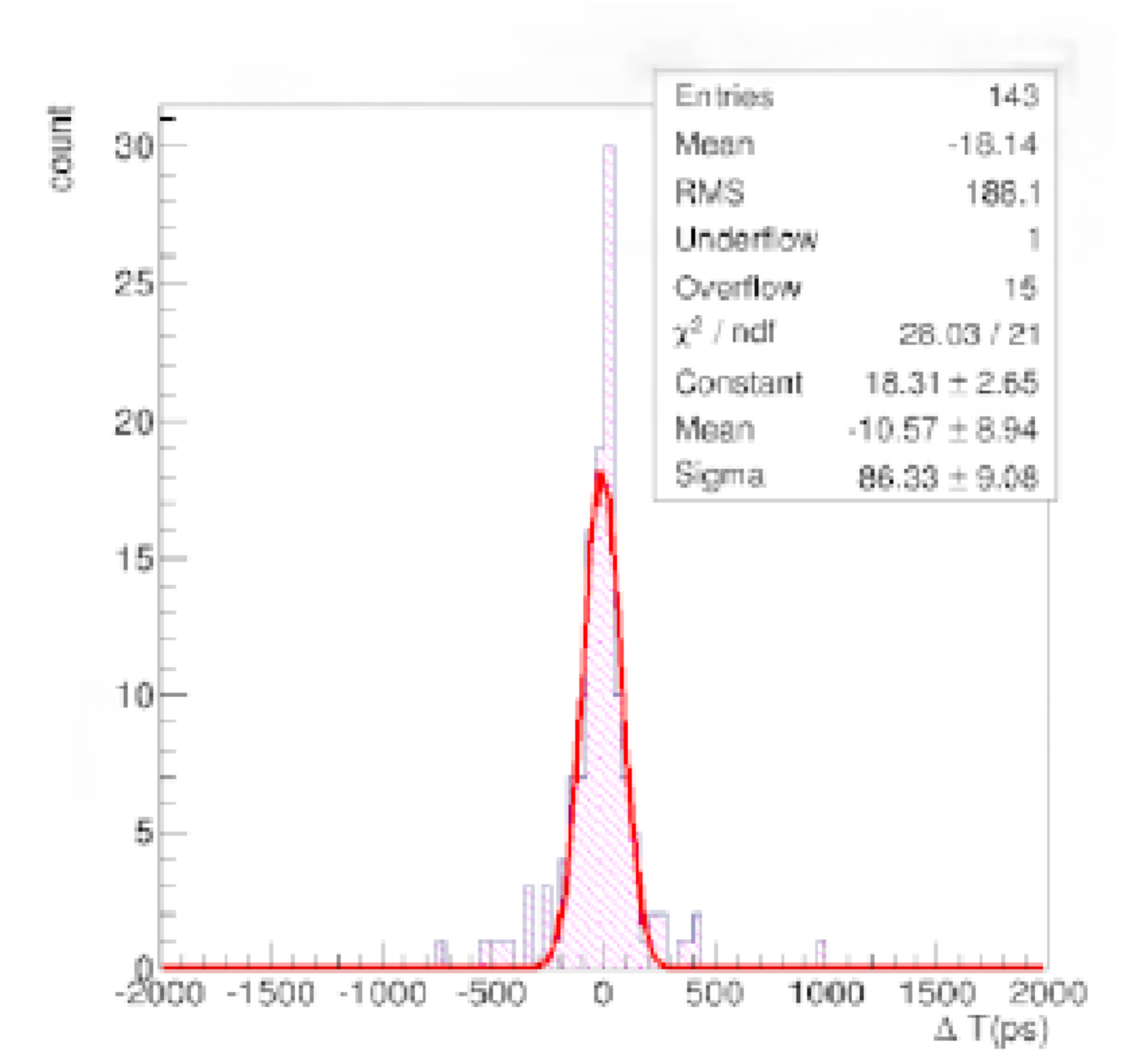}
\caption{MRPC time distribution relative to T$_{0}$ for a pion beam.}
\label{time2}
\end{figure}

The digital timing of the inner and outer MRPCs were filtered by the T1 to T4 scintillators, which required all of them to have signals. The average cluster size of the MRPC was 1.6, as shown in Fig.\ref {cluster}. The timing of inner and outer MRPCs was corrected with respect to Tr0, mainly for the time-amplitude slewing effect. The signal amplitude was estimated by its width (TOT). We have developed a new slewing correction method to correct the relationship between time and amplitude that combines fitting and bin counting. The MRPC timing and TOT plot were divided into several parts. To do the slewing correction, function fitting was used for the parts with enough statistics and bin-by-bin counting method was used for the other parts (Fig.\ref {cluster1}). Compared to the method used in Ref. \cite{yang2014test}, this new method has a better correction effect, especially for the channels with poor statistics. For poor statistics, there is not a suitable function that fits them, so the bin-by-bin method is a good way to do this slewing. The calibration strategy was pad-by-pad or strip-by-strip, so for every single pad or strip, we have very poor statistics. Fig. \ref {time} and \ref {time2} show the distribution of a typical MRPC timing relative to Tr0 for proton and pion beams. The plots show that time resolution of MRPC were measured to be $\sim$160 ps for protons and $\sim$85 ps for pions. It was also found that the time resolution of inner and outer MRPCs are similar. The Figure \ref {eff} shows the efficiency plateau at BEPCII.

\subsubsection{Simulation}
Compared to the beam test result of the MRPCs with similar structure \cite{yong2012prototype,zeballos1996new,shao2008upgrade,hu2017t0,akindinov2004results}, the time resolution obtained for the simulation was significantly worse. The reason for this discrepancy is that, for beam test described in Ref. \cite{yang2014test}, the MRPCs were located in the middle position of the Tr0 detector. Thus the formula Tr0=$\frac{(T1+T2+T3+T4)}{4}$ provides a good estimation of the reference time for the MRPCs. However, in this work, the inner and outer MRPCs were placed downstream of the beam line at distances of $\sim$80 and $\sim$100 cm, respectively, from the geometric centre of the Tr0 detector. Because of the beam momentum uncertainty and the interaction with the detector materials that led to energy loss, multi-scattering etc., there was additional timing jitter compared to the measurement in Ref. \cite{yang2014test}. This means what we measured and reported in Fig. \ref {time} and \ref {time2} are actually $\Delta T$=$T_{MRPC}-(Tr0+T_F)$, where T$_F$ is the time of flight between the Tr0 and CSR-T0 MRPC modules. T$_F$ varies because of beam momentum variation, so the intrinsic MRPC time resolution with this effect taken into account should be $\sigma_{MRPC}$=$\sqrt{\sigma^2_{\Delta T}-\sigma^2_{Tr0}-\sigma^2_{T_F}}$.\par
To quantitatively understand the experiment results, we used the GEANT4 toolkit \cite{agostinelli2003geant4} to simulate the beam test experiment at IHEP-E3. The beam test system was simplified in the simulation by only considering the most relevant detectors, including the two thin slices of plastic scintillator (SC1 and SC2, 5 cm x 5 cm x 0.5 cm) used for beam trigger, two groups of plastic scintillator strips (T1/T2 and T3/T4, 5 cm x 2 cm x 1 cm) used as the Tr0 detector, four CBM-TOF MRPC modules, and two CSR-T0 MRPC modules arranged along the beam direction (See Fig.\ref{setup}). As much as possible, we used the materials used in the actual situation in the Geant4 description. Along the beam direction, each MRPC module mainly includes the following materials: an aluminum gas-tight shielding box (2-4 mm thick in total), three pieces of PCB (3-6 mm thick in total), 12 pieces of glass plate (6 mm thick in total), and a gas sensitive region (2.5 mm thick in total).\par
\begin{figure}[htb]
\centering
\includegraphics*[width=60mm]{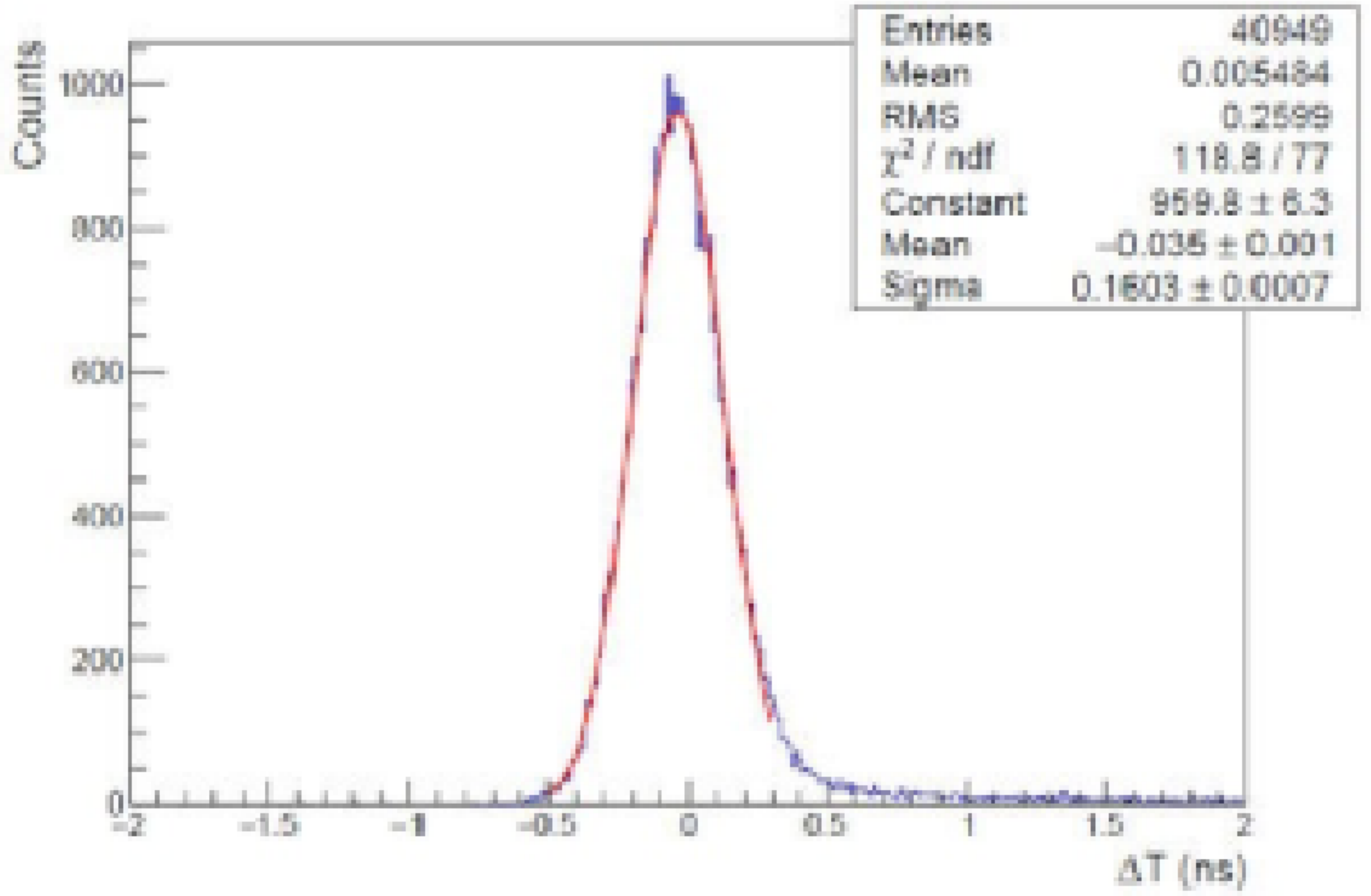}
\caption{Simulated distribution of T$_{MRPC}$ - Tr0 with proton beam. The beam momentum uncertainty and Tr0 timing jitter are included.}
\label{time3}
\end{figure}

The beam momentum resolution was measured to be $\sim$2.5\% at IHEP-E3 at an injection hadron beam momentum of 700 MeV/c. These parameters were considered in the GEANT4 simulation. The energy loss and the multiple Coulomb scattering were also taken into account. The reference time Tr0, was the the average of all four channels of the Tr0 detector. An intrinsic MRPC timing uncertainty of 40 ps and a Tr0 timing uncertainty of 20 ps were smeared into the simulation data. Fig \ref {time3} shows the distribution of T$_{MRPC}$ - Tr0. By comparing Fig.\ref {time3} to Fig.\ref {time}, we see that the simulation and experimental results are consistent with each other, indicating that the MRPC time resolution is 40ps. Note that the Tr0 time jitter is smaller in the simulation (20 ps) than that shown in experiment (40 ps, Fig. \ref{distiguish}), which includes an additional contribution from beam momentum uncertainty as well as the intrinsic Tr0 uncertainty.\par
\begin{figure}[htb]
\centering
\includegraphics*[width=60mm]{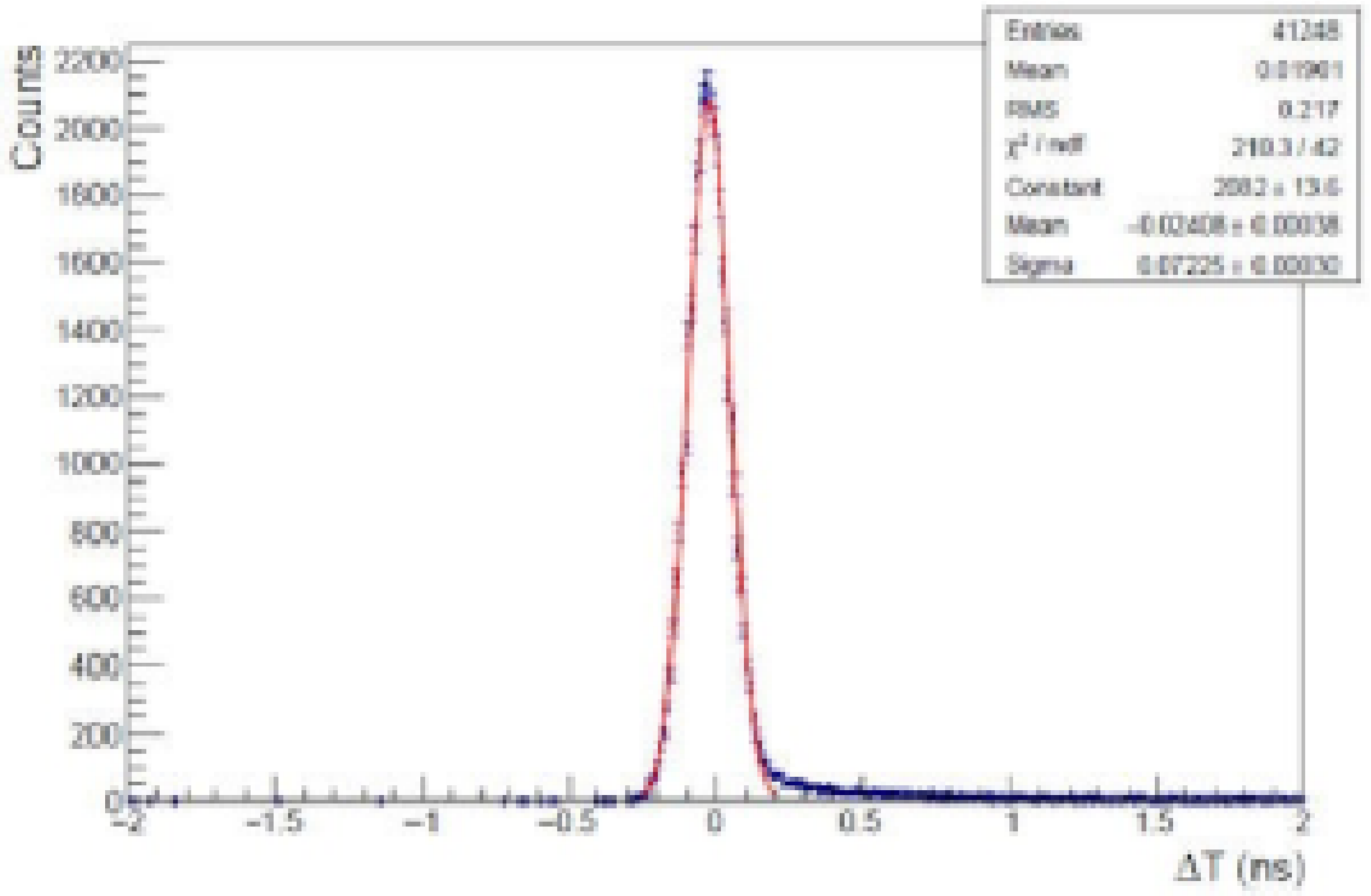}
\caption{Simulated distribution of T$_{MRPC}$ - Tr0 with pion beam. The beam momentum uncertainty and Tr0 timing jitter are included.}
\label{time4}
\end{figure}

The above comparison was done for a proton beam. Because the IHEP-E3 beam also contains a small fraction of pions, we also compared the GEANT4 simulation results to the experimental MRPC time response for pions. We found that with an MRPC intrinsic timing uncertainty of 50ps and a Tr0 intrinsic timing uncertainty of 40ps smeared into the simulation, the simulated distribution of T$_{MRPC}$ - Tr0  (shown in Fig.\ref {time4}) is consistent with the experimental results. The long tail on the time spectra is caused by the Coulomb multiply scattering and dE/dx effects of the beam, which are more significant for protons than for pions. \par
Through these analyses of simulation and experiment data, we conclude that the time resolution was $\sim$40 ps for proton and $\sim$50 ps for pion at 700 MeV/c, for both inner and outer MRPC modules. These values are also consistent with the results from previous test with similar MRPC structure.  \cite{yong2012prototype,hu2017t0,besiii2009construction,yang2014test}. 

\section{Heavy-ion in beam test}

\subsection{Experimental setup}
\begin{figure}[htb]
\centering
\includegraphics*[width=80mm]{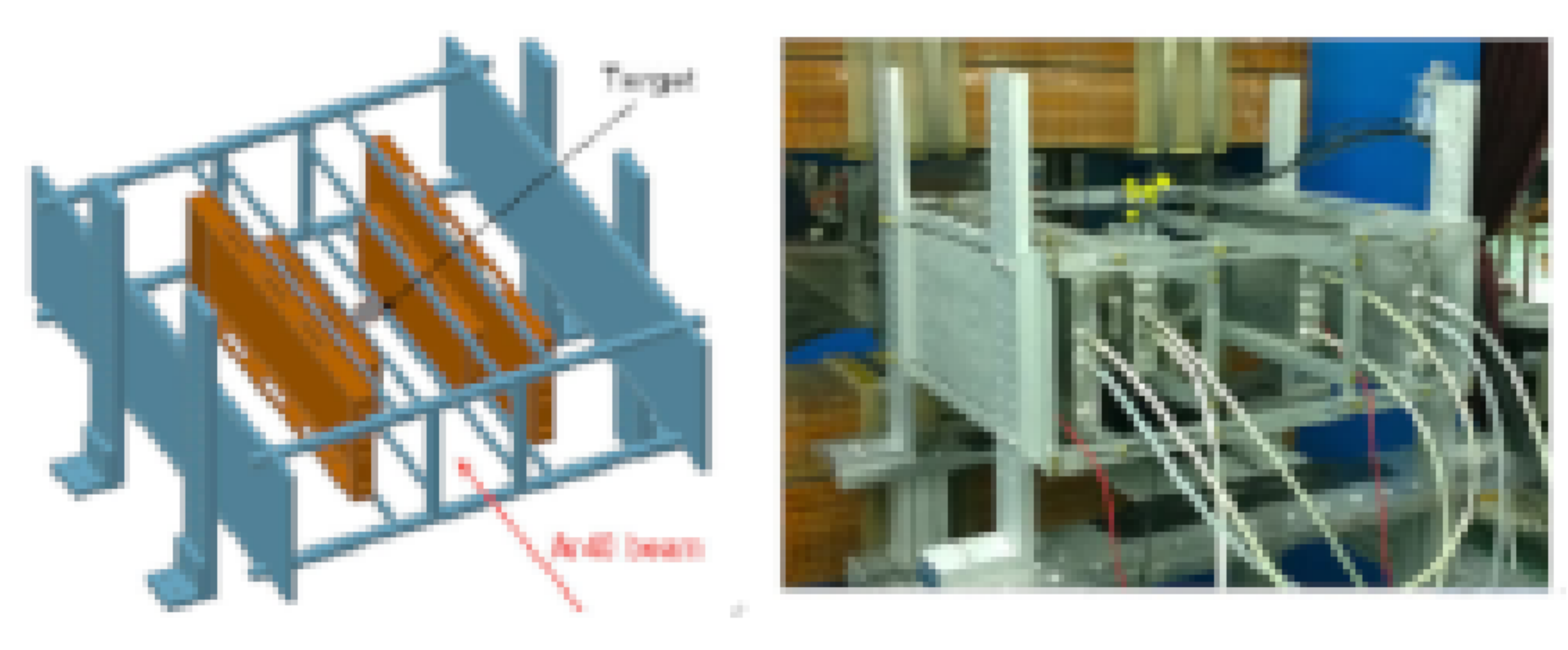}
\caption{Design of the T0 detector structure (a) for the beam test at CSR and a photograph of the experiment(b).}
\label{heavysetup}
\end{figure}

In November 2016, the prototype of the T0 detector was tested with the heavy-ion beam at CSR . Fig. \ref {heavysetup} shows a sketch of the T0 detector test setup for the CSR external-target experiment. Only 1/4 of the full T0 detector was built and tested, including two inner and two outer MRPCs. A photograph of the MRPC modules and mechanical structure is also shown in Fig. \ref {heavysetup}.\par
An Ar-40 beam with a kinetic energy of 300 AMeV bombarded a lead and carbon target that was located at the geometrical centre of the T0 detector. Operated in a stand-alone mode, the T0 system was self-triggered by requiring all four MRPCs be fired. The MRPC signal, after amplification and discrimination by the FEE, was sent to the digitalization electronics via a long cable ($\sim$10 m). The total number of readout channels was 80, but we did not have enough electronics for this. Therefore, we used two kinds of TDCs with 40 channels were recorded by FPGA TDC modules and the other 40 channels were processed by the previously designed time digitization modules based on HPTDCs. Synchronization between these two types of TDC modules was achieved based on two techniques. First, a 40 MHz system clock was fed to all the TDC modules, and thus the coarse time and fine time (after interpolation) were all synchronized with the phase of this clock signal. Second, after the electronics are powered up, a global reset signal is generated, and it was fanned out from the sub-trigger module to all the TDC modules to clear the coarse time counter value and align the “start” point for time measurement. \par
\begin{figure}[htb]
\centering
\includegraphics*[width=75mm]{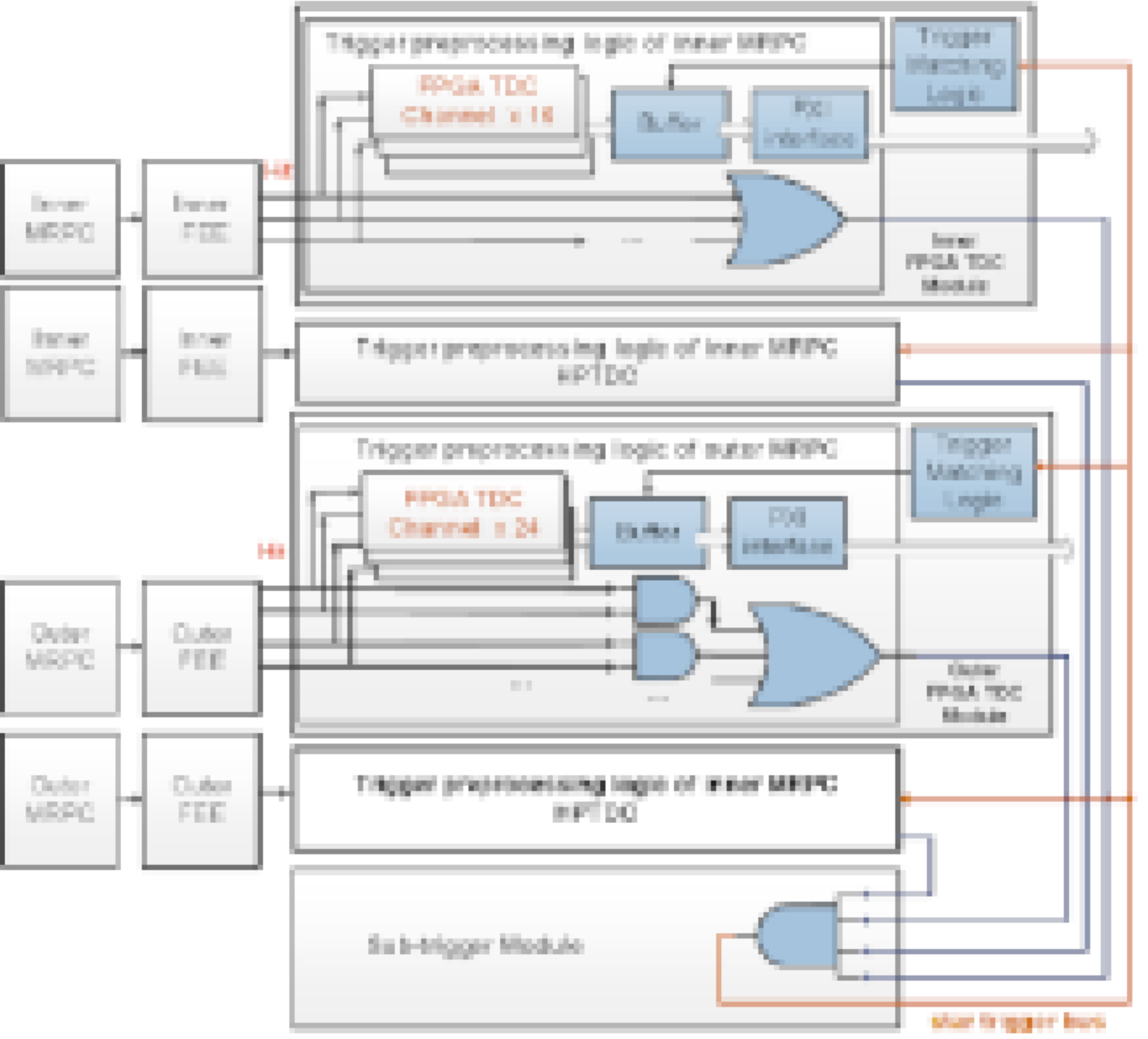}
\caption{Trigger processing and data readout.}
\label{system}
\end{figure}

Fig.\ref {system} shows that the data from FPGA TDCs were stored in the internal buffer inside the FPGA, and the valid data were read out when a trigger signal was received. Trigger processing was organized in two hierarchies.  The trigger mode in this experiment was as follows. In the first step, for the inner MRPC, the hit signals were fed to an logic OR gate. For the outer MRPC, the hit signals from the two ends of one MPRC strip were input to an logic AND gate and then further processed by the following logic OR gate. The above processing functions were implemented in the TDC modules. Next, the flag signals from both the inner and outer MRPC electronics modules were sent to the sub-trigger module and processed by its logic AND gate. Finally, a trigger signal was generated and transmitted to all TDC modules through the star trigger bus in the PXI crate. 
\subsection{Heavy-ion in-beam test}
\subsubsection{HV scan}
\begin{figure}[htb]
\centering
\includegraphics*[width=60mm]{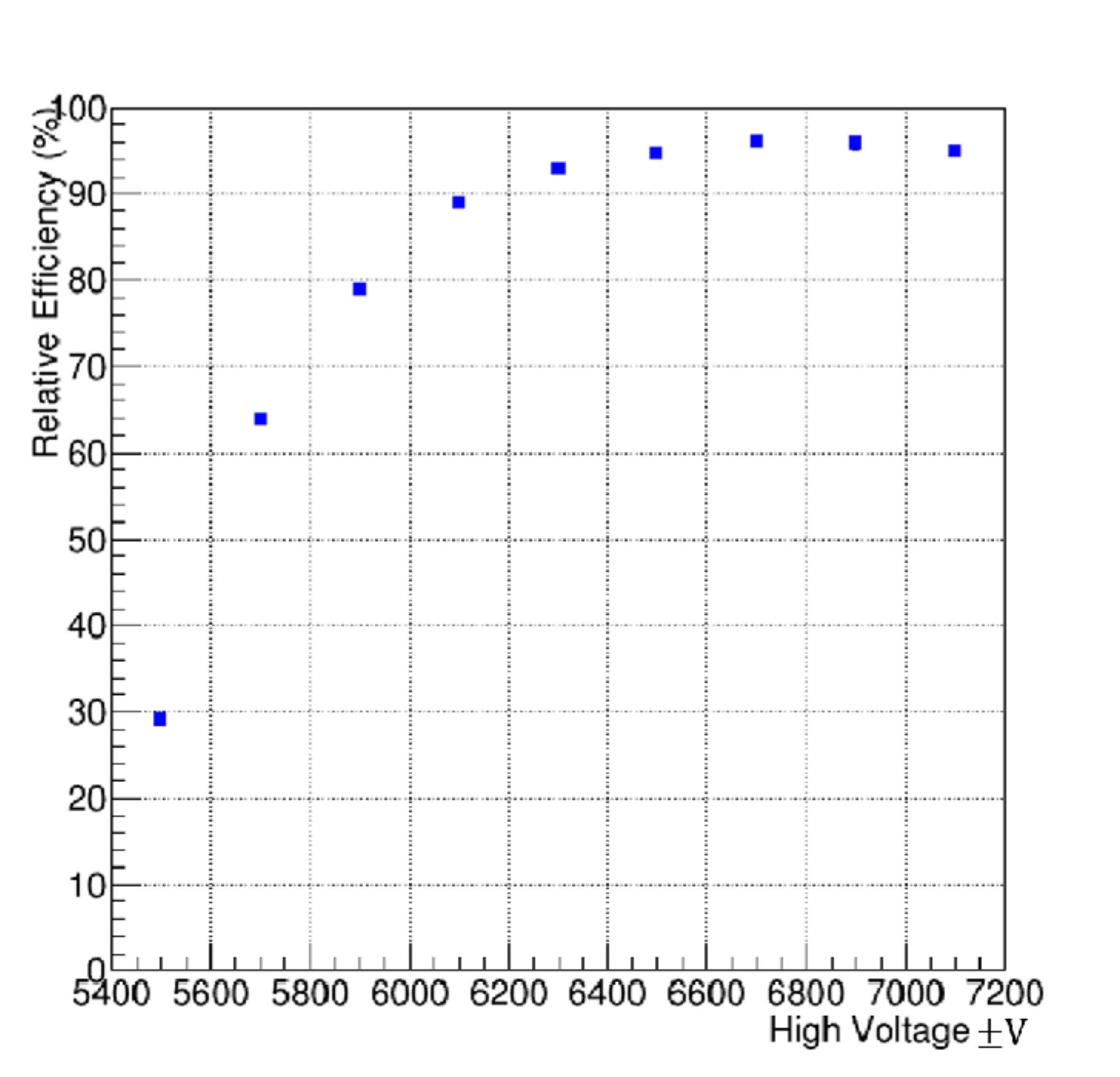}
\caption{MRPC detection efficiency vs. HV at CSR\cite{hu2017t0}.}
\label{hv}
\end{figure}

According to the rule V =$V_a\frac{P_0T}{PT_0}$  \cite{gonzalez2005effect}, where $V_a$is the applied voltage, T represents the operating temperature and P denotes the gas pressure, the operating voltage V changes with P, and thus with altitude. Therefore, The operation condition of MRPC must adapt to the change of altitude in Lanzhou area ($\sim$1500 m a.s.l,with a normal atmosphere pressure of 5/6 bar). We set up a cosmic-ray test during the beam time at CSR and checked the detection efficiency as a function of the applied HV, See Fig. \ref{hv}\cite {hu2017t0}. The efficiency was a relative efficiency without correction for the acceptance of cosmic-ray, but one can clearly see a plateau. The working HV was chosen to be 6800V, which is significantly lower than the normal HV ($\sim$7200V) for tests at the IHEP-E3 line (Fig.\ref{eff}) and in the laboratory.  
\subsubsection{Calibration procedure}
During the beam test the system was self-triggered, and there was no reference time and tracking information for the T0 detector. The event vertex position and field map that electronically channel match with the pad or strip number were also lacking. Therefore, the basic calibration strategy was to perform a relative correction to each channel. The time offset, TOT slewing correction, and particle velocity correction all needed to be calibrated.\par
\begin{figure}[htb]
\centering
\includegraphics*[width=80mm]{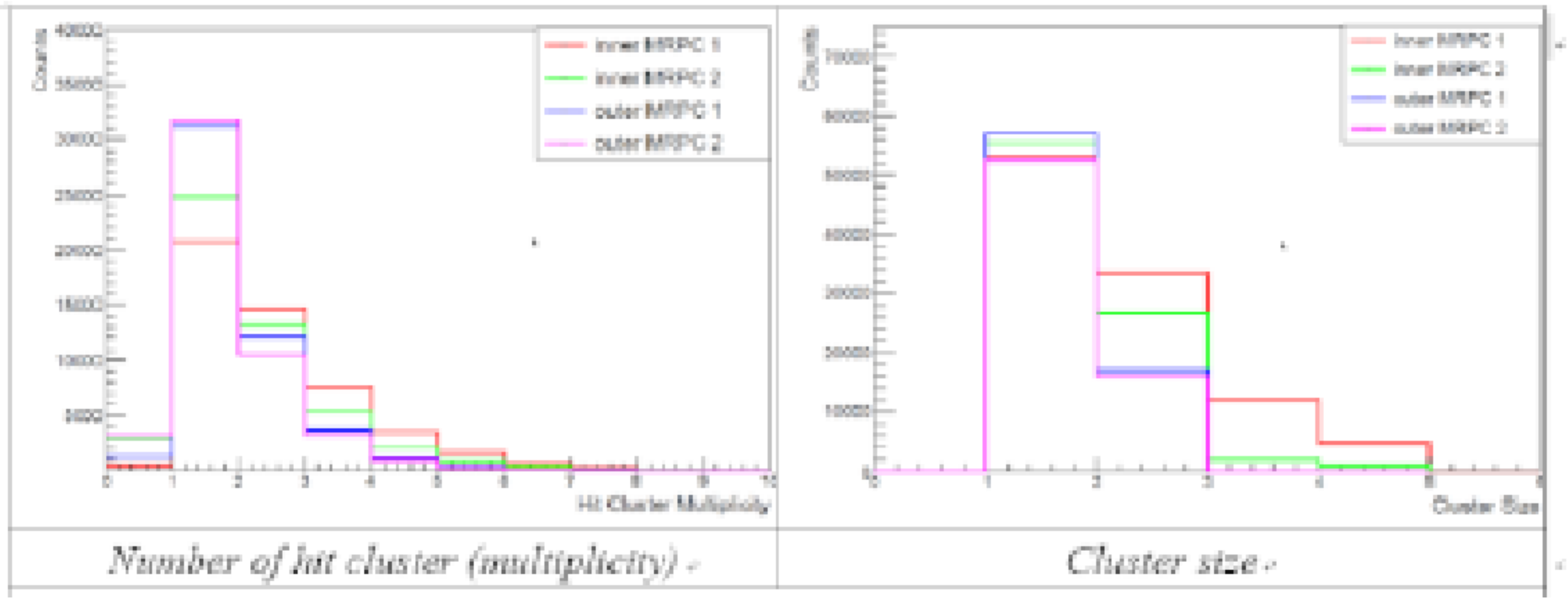}
\caption{Number of hit cluster and cluster size of the MRPC at beam time.}
\label{clusterheavy}
\end{figure}
\begin{figure}[htb]
\centering
\includegraphics*[width=70mm]{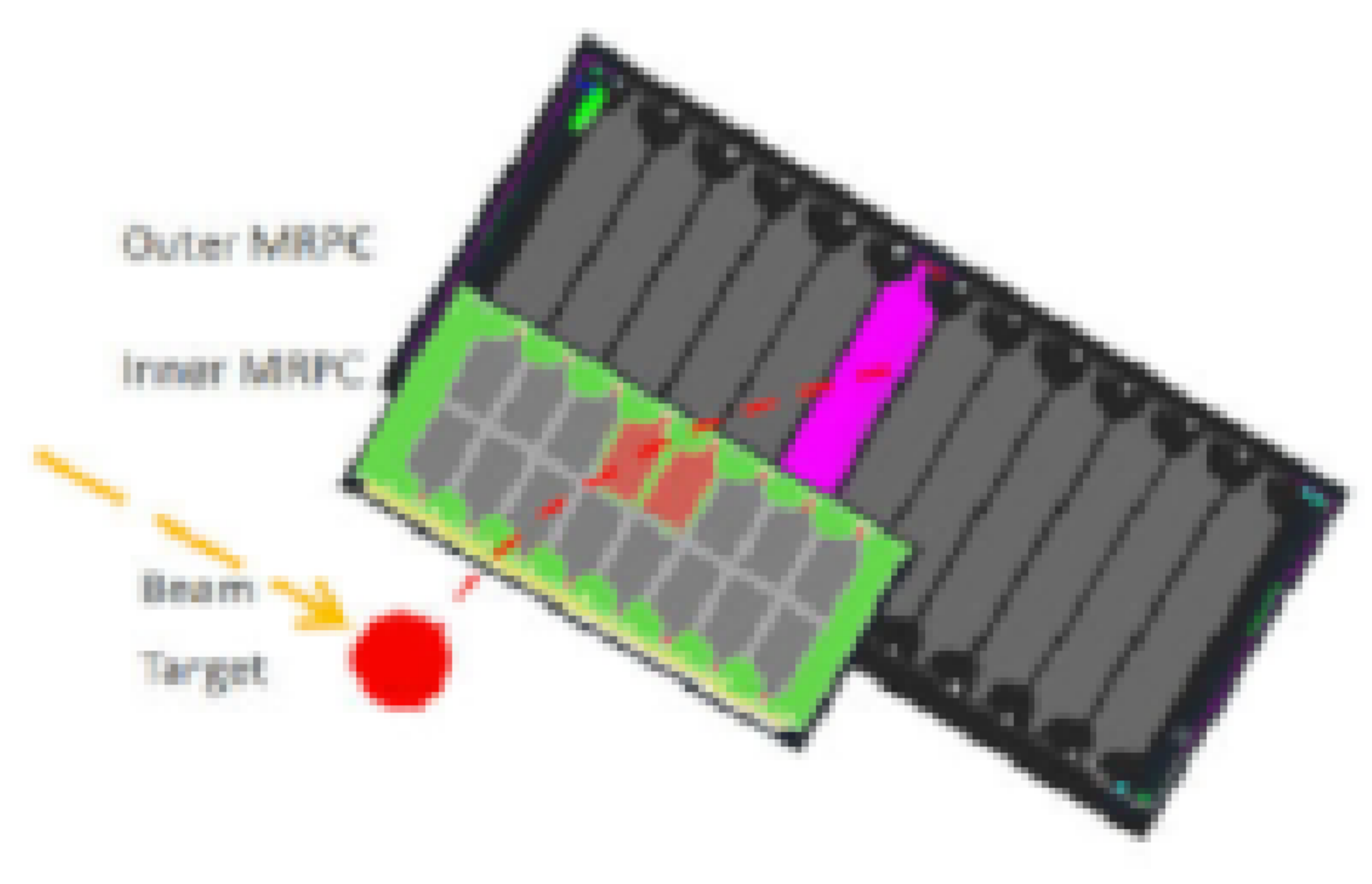}
\caption{Event selection: Two neighbouring pads of inner MRPC and one strip of the outer MRPC were fired by a single particle.}
\label{strategy}
\end{figure}

The first step was to tune the time offset of each channel by comparing signals from neighbouring pads fired by a single particle. Each pair of inner and outer MRPCs of the T0 detector was combined to form one group (two groups in total in our test), and each group was calibrated separately. Fig.\ref {clusterheavy} shows the cluster size and hit multiplicity plots. The maximum cluster size of the inner MRPCs (pad readout) was four, while for outer MRPCs (strip readout) was two. To suppress background hits, one fired strip on outer MRPC and two fired pads on inner MRPC within each group were required when calibrating the inner MRPC module. Fig.\ref {strategy} shows that the selected particle first hits the inner MRPC and then hits the outer MRPC, firing two neighbouring pads, so the hit point should be near the boundary region between the pads. In this case, when a single particle passes, it causes two fired channels of the inner MRPC, so their hit time should be the same. If channel 0  is considered to be the reference, and if all channels are iterated, the relative time offset can be evaluated and calibrated by a simple time shift for each channel. The calibration procedure for the outer MRPC modules is similar. In this case, it is necessary for there to be only one fired channel on both inner and outer MRPCs.\par
\begin{figure}[htb]
\centering
\includegraphics*[width=60mm]{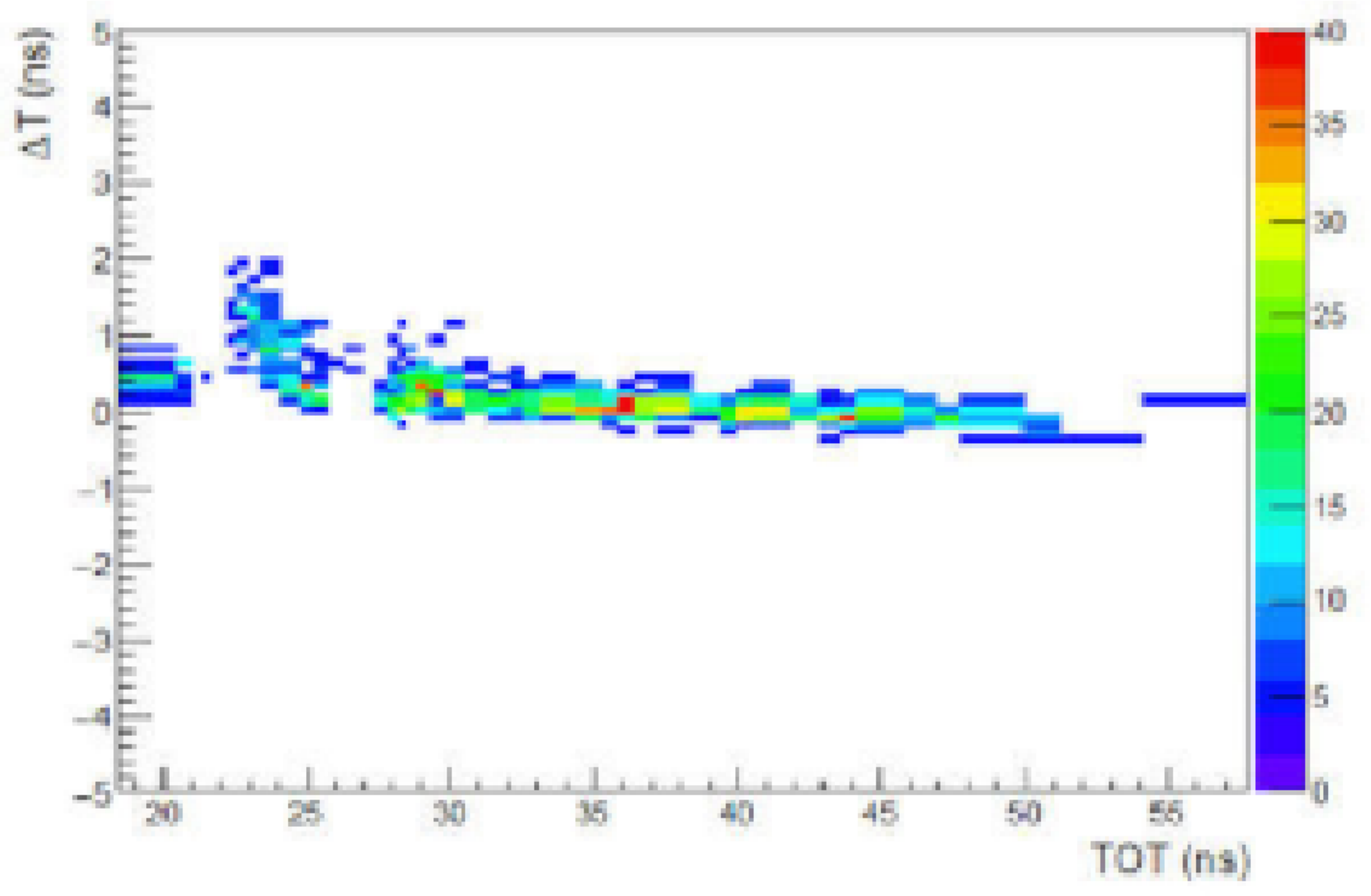}
\caption{Typical MRPC time vs TOT slewing correlation.}
\label{ta}
\end{figure}

The second step is to correct the time slewing effect for each channel. As in the step 1 calibration, a single charged particle was fired at both inner and outer MRPCs. Next, the time difference between neighbouring channels was plotted as a function of TOT$_i$ , where TOT$_i$ is the measured signal width of the channel to be calibrated. In our analysis the neighbouring pads/strips were those directly sharing one boundary with the selected pad/strip. For the inner MRPCs, one pad had three neighbouring pads (two neighbours for pads at the ends), and for outer MRPCs, one strip had two neighbouring strips (one neighbour for strips at the ends). For the outer MRPCs, TOT$_i$ was the mean TOT measurements from both ends. A typical slewing effect is illustrated in Fig.\ref {ta}. A table of bin-by-bin centre value, rather than a fit curve, was used to correct the TOT dependence. The procedure was repeated until convergence was observed.\par
\begin{figure}[htb]
\centering
\includegraphics*[width=60mm]{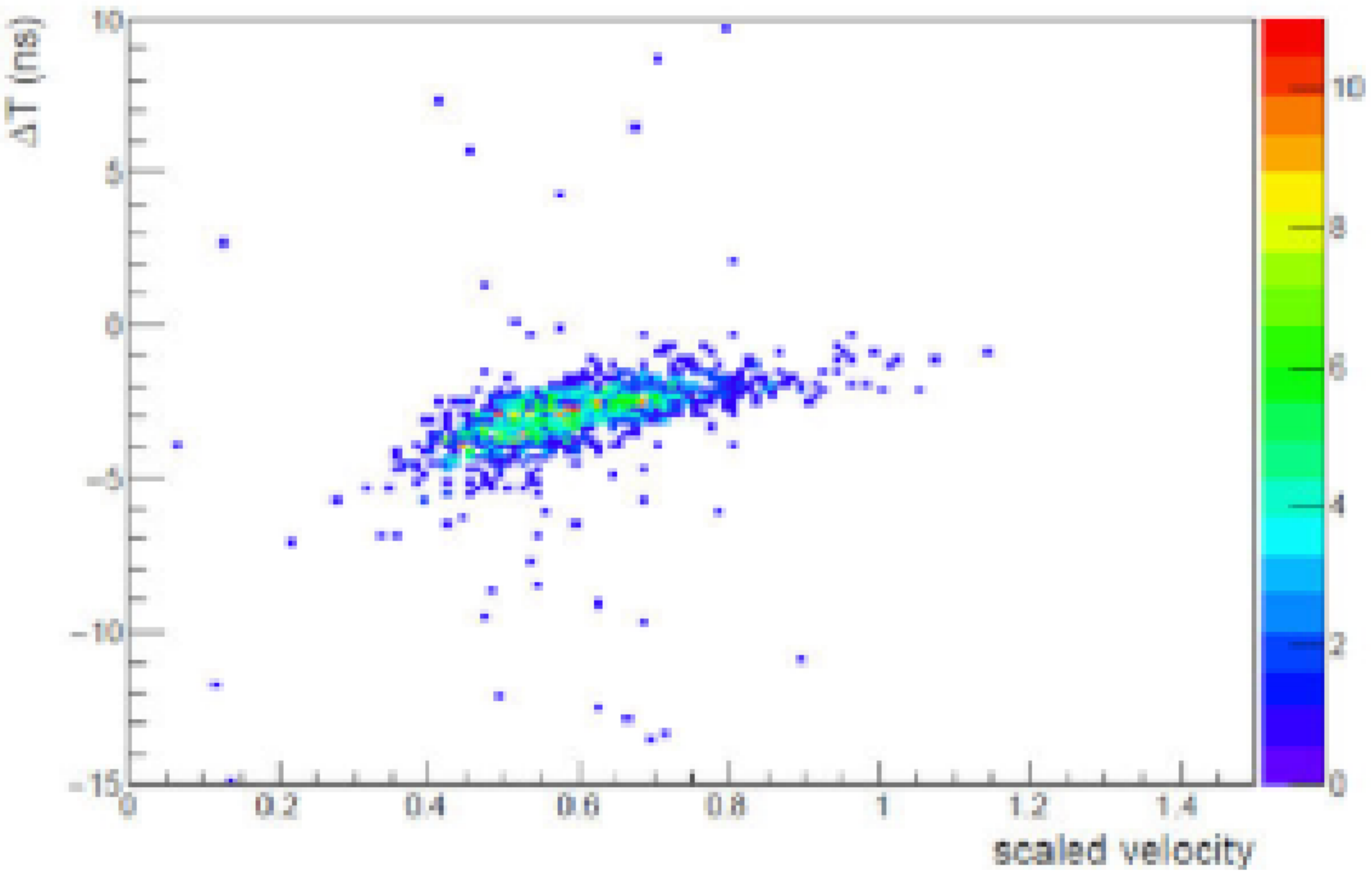}
\caption{Typical start time difference between two T0 groups vs. effective particle velocity.}
\label{tv}
\end{figure}

The next step in the calibration concerns the particle momentum spread. In the beam test, the momenta of the final state charged particles varied significantly from $\sim$200 to 600 MeV/c. Particles with momenta <200 MeV/c are likely to be absorbed or scattered by the detector materials. The particle speed can be estimated by the time difference between the inner and outer MRPCs in the same group and their distance, v = $\frac{L_{Out}-L_{In}}{T_{Out}-T_{In}}$ , where $L_{Out}$ ,$L_{In}$ ,$T_{Out}$  and $T_{In}$ are the flight lengths and times from the collision point to the outer and inner MRPC. For this test $L_{Out}$ =22.5 cm and $ L_{In}$ = 12.5 cm. The event start time, T0 , can be calculated by T$_0$ =$\frac{T_{In}L_{Out}-T_{Out}L_{In}}{L_{Out}-L_{In}}$  (Eq. 1). This formula provides accurate collision time if there is no energy loss and if no multiple scattering effects are involved. However, at CSR energy, these factors cannot be ignored. Because relevant timing measurement from both groups of the MRPCs was necessary, to do a velocity calibration, we needed a reference Tr0. This was done by requiring each of the two groups of MRPCs to contain at least one valid track hitting both inner and outer MRPCs. Thus each group of MRPCs can give a measurement of Tr0, which can be used as a (relative) reference for the other group. The $T_0$ difference between the two groups was plotted vs. particle speed, as shown in Fig.\ref {tv} \cite{hu2017t0}. A clear velocity dependence was seen and used to calibrate the value of $T_0$. \par
There are some other factors that should be noted, such as the magnetic field and collision vertex uncertainty. The magnetic field was found to be <0.1 Tesla at the T0 detector location, so the bending radius of a proton was >6.7 m if the momentum was required to be >200 MeV/c. Compared to the flight length $L_{Out}$ and $L_{In}$, the effect of magnetic field is small and, therefore, the effect was neglected in this analysis. The heavy-ion beam had a round shape and a root mean square (RMS) radius of 3 mm. Because there was no measurement of the collision vertex position, this uncertainty also affected the T0 detector time resolution.
\subsubsection{Time resolution}
\begin{figure}[htbp]
\centering
\begin{minipage}[t]{1.0\linewidth}
\centering
\includegraphics*[width=60mm]{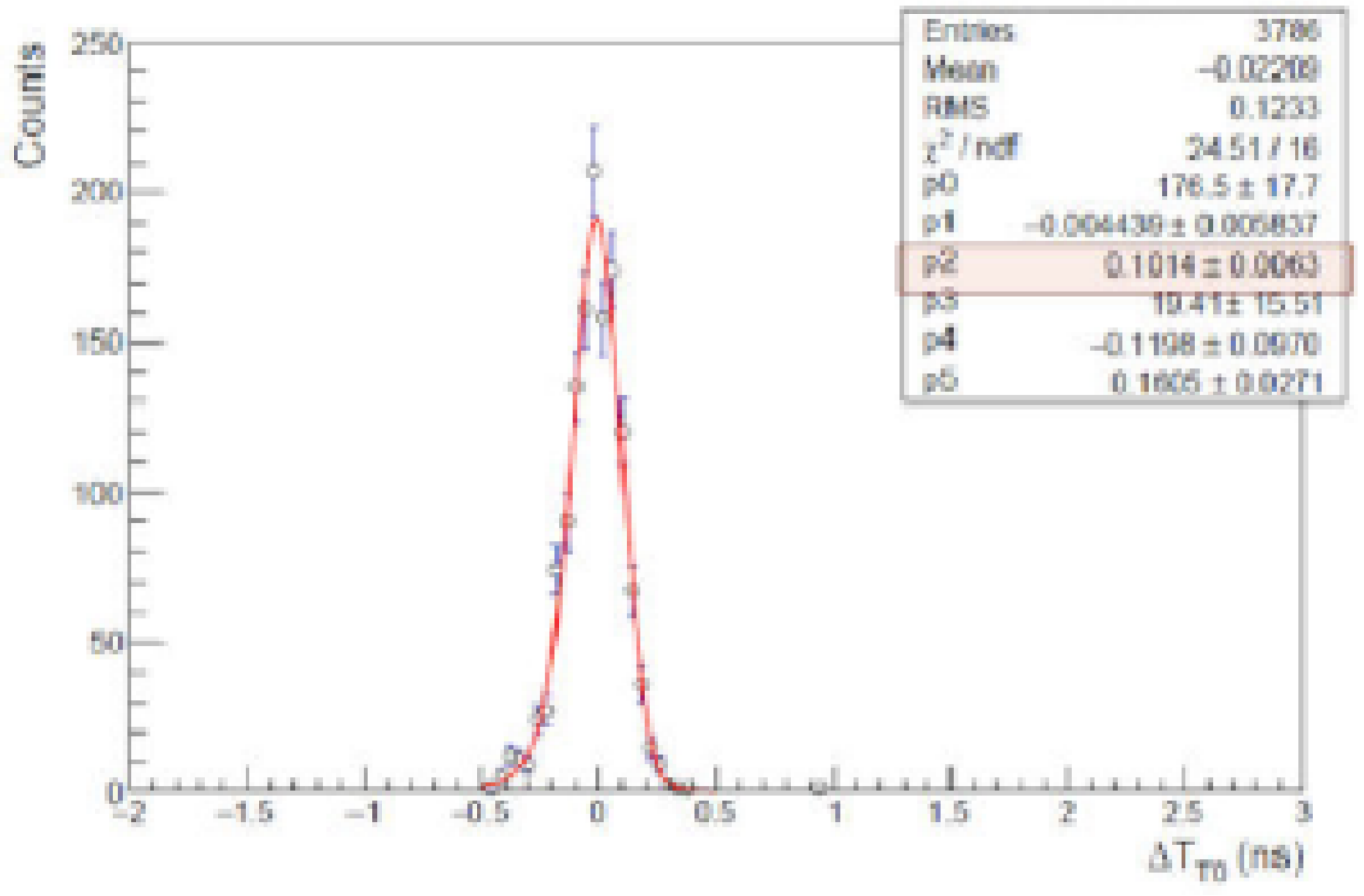}
\centerline{(a)}
\end{minipage}
\vfill
\begin{minipage}[t]{1.0\linewidth}
\centering
\includegraphics*[width=60mm]{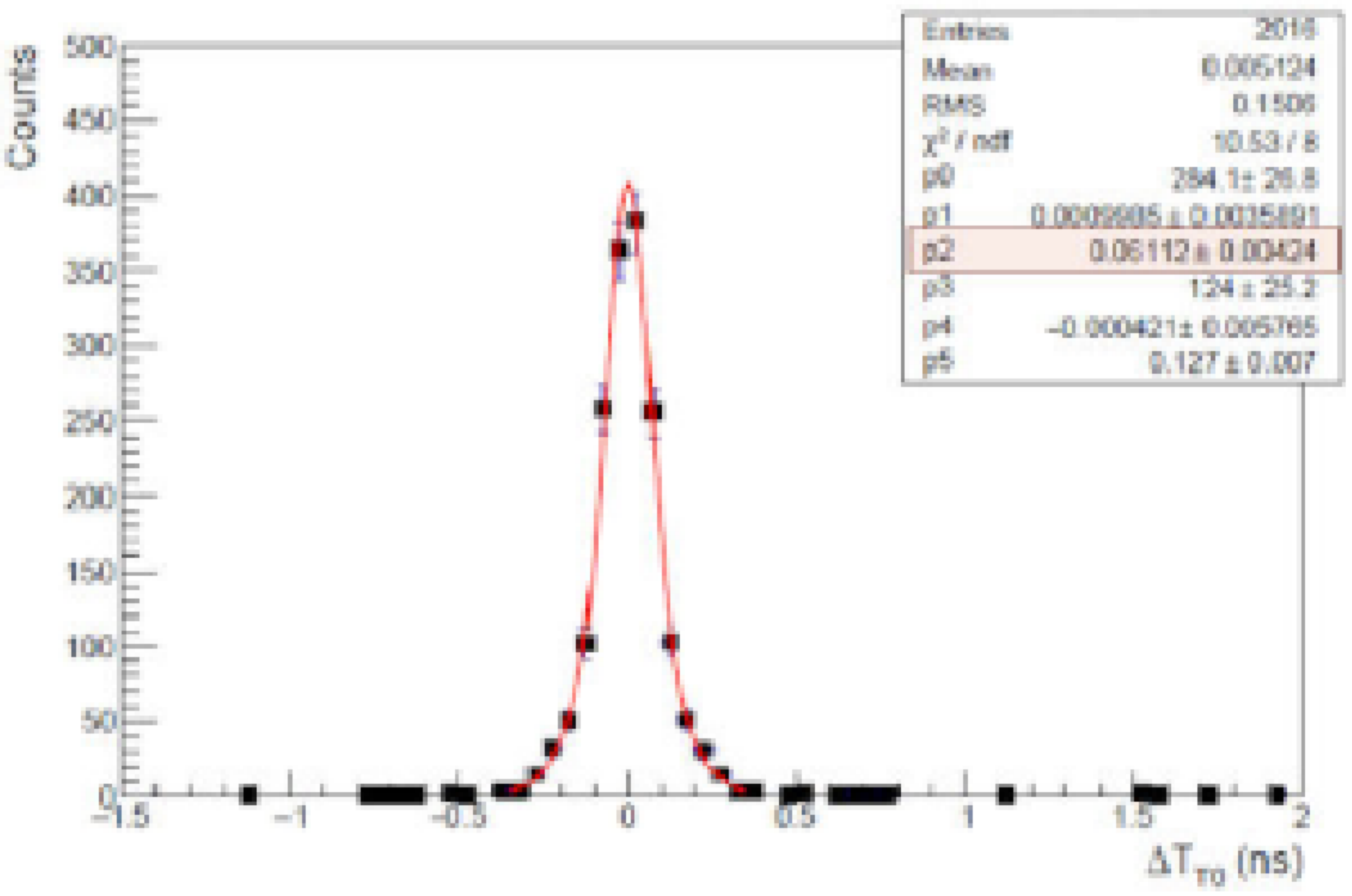}
\centerline{(b)}
\end{minipage}
\caption{Start time difference between two T0 groups (top), and within one T0 group (bottom).}
\label{time6}
\end{figure}

\begin{figure}[htbp]
\centering
\begin{minipage}[t]{1.0\linewidth}
\centering
\includegraphics*[width=60mm]{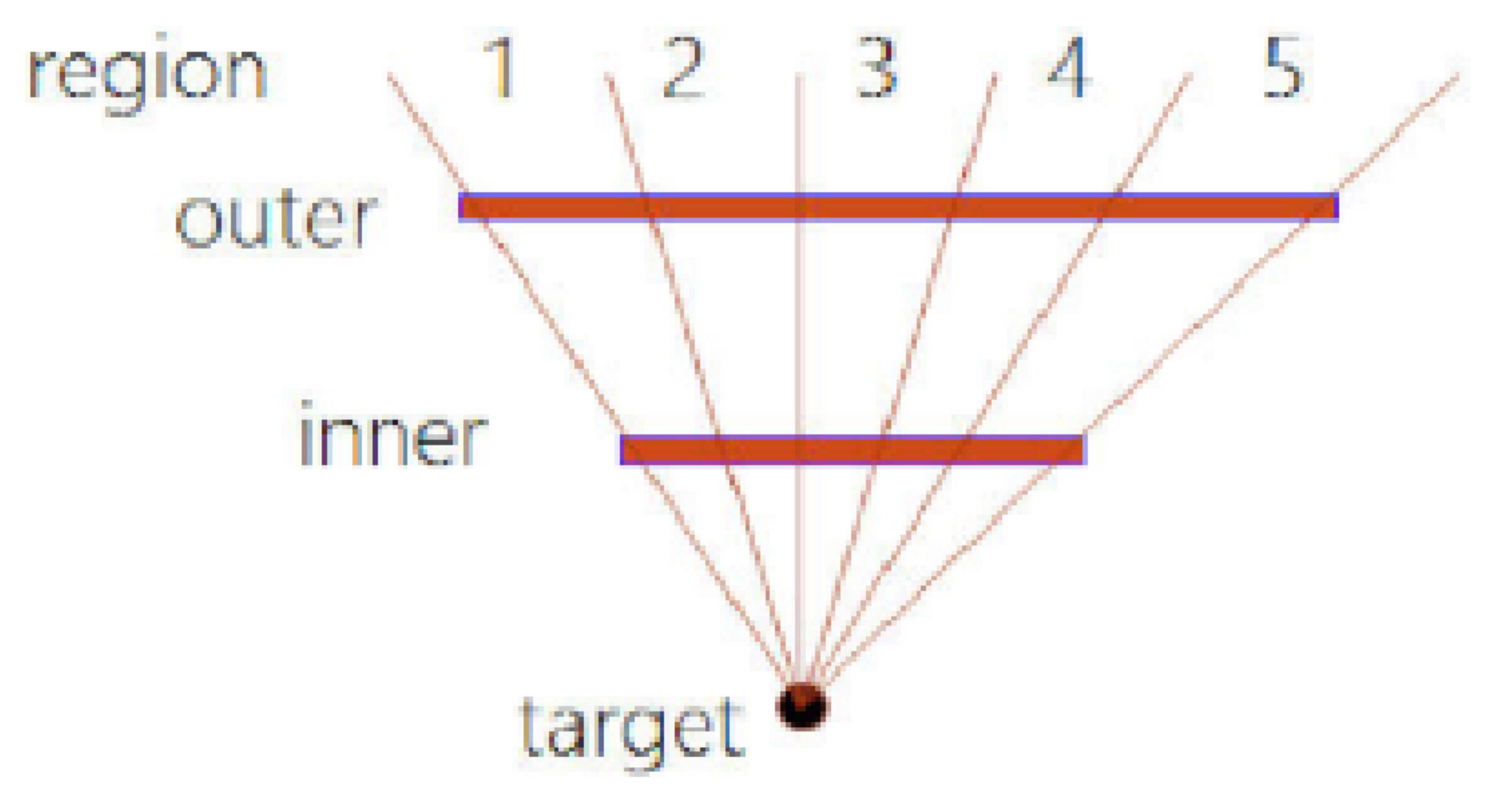}
\centerline{(a)}
\end{minipage}
\vfill
\begin{minipage}[t]{1.0\linewidth}
\centering
\includegraphics*[width=60mm]{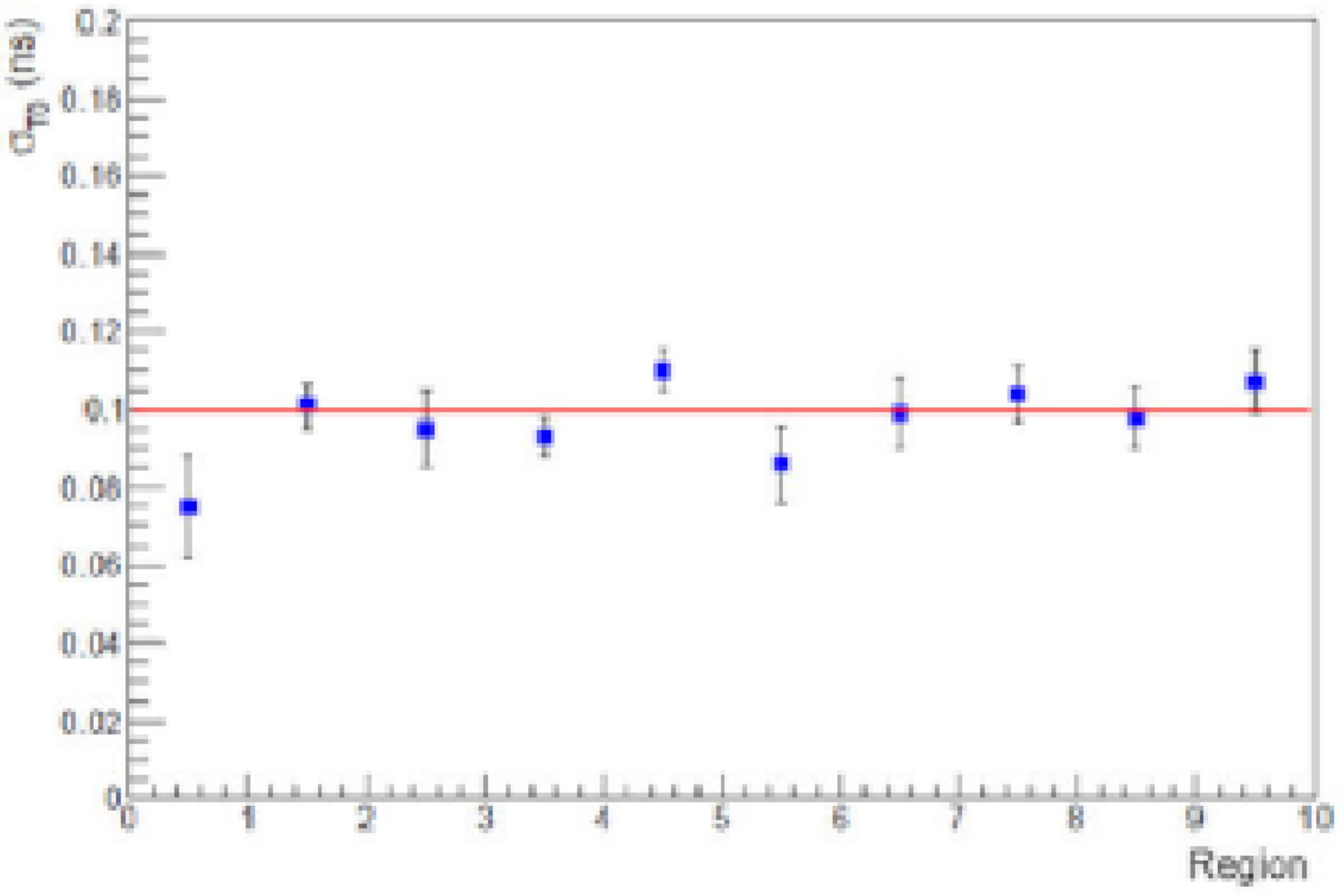}
\centerline{(b)}
\end{minipage}
\caption{(Top) Different hit position regions along the beam direction. (Bottom) T0 time resolution in each region.}
\label{hitposition}
\end{figure}

According to Eq. 1, the T0 detector's time resolution is mainly determined by the MRPC timing accuracy, the particle flight length, and momentum spread. The vertex uncertainty also affects the resolution by changing the flight length. To estimate the timing performance of the T0 detector, the start time differences between the two T0 groups are calculated using $\Delta T_{T0}$ =$\frac{T_{01}-T_{02}}{2}$. Fig. \ref {time6} (top a) shows their distribution after all corrections were applied. In this plot, each group was required to be hit by only one track, so $\sigma_{\Delta_{T0}}$ represents a good measure of the T0 detector resolution, $\sigma_{T0}$, by assuming $T_0$ =$\frac{T_{01}+T_{02}}{2}$. When there were two tracks recorded by the T0 detector(one track for each group), the T0 time resolution was found by double-Gaussian fitting to be $\sim$100 ps. See Fig. \ref {time6} (top a).
We further studied the response uniformity of the T0 detector. Each group of the detector was divided into five regions according to the hit position along the beam direction. A total 10 regions were scanned, and the time resolution was measured in a similar way to what was done to generate Fig. \ref {time6}. The result is shown in Fig. \ref {hitposition} (bottom b). It is clear that a uniform performance of the MRPCs was achieved. \par
Besides MRPC timing uncertainty, the observed T0 time resolution of $\sim$100 ps (Fig.\ref{time6} (a) and (b)) included contributions mainly from the collision vertex uncertainty, which was measured to be $\sigma_{VTX}$= 3 mm in the plane perpendicular to the beam direction. To study this contribution to the uncertainty, events with two or more tracks hitting one of the two groups were selected, and the $\Delta T_{T0}$ distribution was drawn for each pair of tracks. Because both tracks were from the same group, the effect of the vertex position variation largely cancelled out. With a double-Gaussian fit, the T0 time resolution of $\sim$60 ps was determined. It is worthy noted that, in proposed CEE operation, the collision vertex will be precisely measured by other detectors, so it should not contribute to T0 time resolution.
Both Ar+C and Ar+Pb collision data are analyzed. The results were found to be very similar and consistent with each other, despite some differences in hit multiplicity.

\subsubsection{Simulation study}
\begin{figure}[htbp]
\centering
\begin{minipage}[t]{1.0\linewidth}
\centering
\includegraphics*[width=60mm]{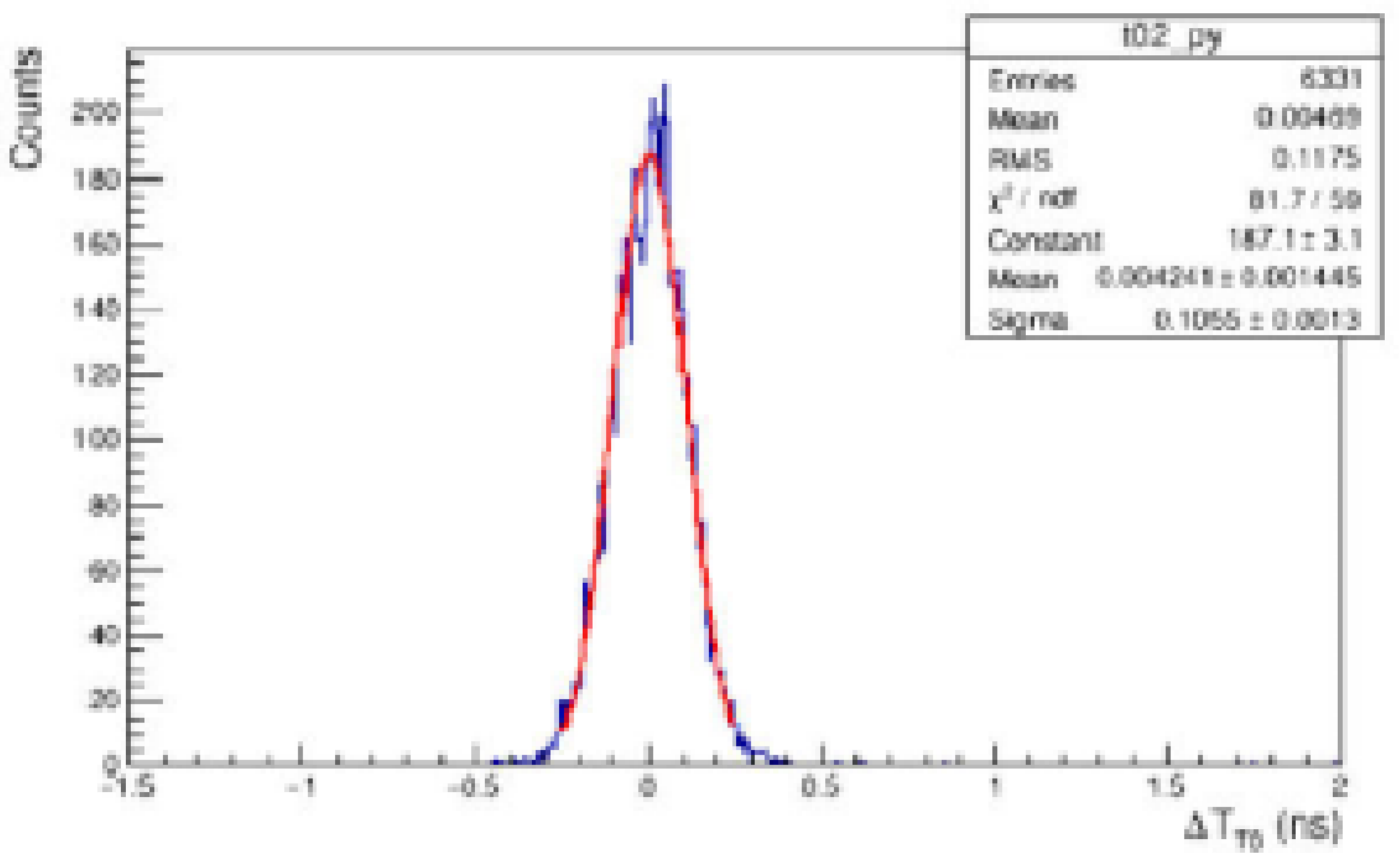}
\centerline{(a)}
\end{minipage}
\vfill
\begin{minipage}[t]{1.0\linewidth}
\centering
\includegraphics*[width=60mm]{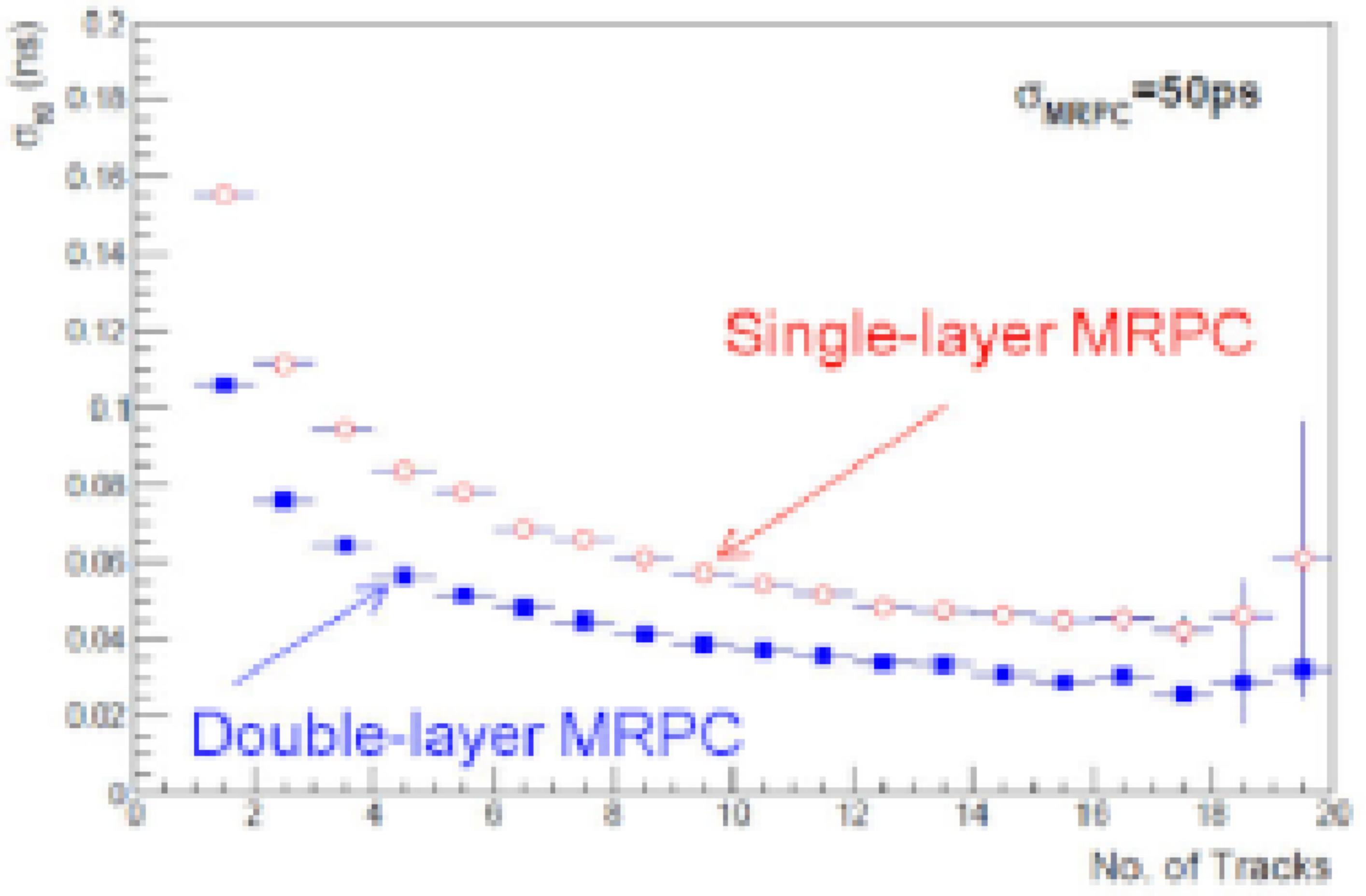}
\centerline{(b)}
\end{minipage}
\caption{T0 time resolution from simulation data.}
\label{simulation}
\end{figure}

Owing to the lack of reference information of the collision time and other properties of the final state particles, we used an MC simulation to model the experimental result and provide a performance expectation for the T0 detector. The GEANT4 toolkit was used for the description of the detector and its response to particles generated by heavy-ion reaction. Only the MRPCs, along with the boxes and FEE (the brown parts in Fig. \ref {heavysetup}) were included in the GEANT4 simulation. The target and other supporting structure were ignored. To improve the simulation efficiency, the T0 detector was assumed to have full acceptance, like in Fig.\ref {magnet}. The heavy-ion collision event was simulated by the UrQMD3.4 generator \cite{A:2005}. The incident beam consisted of argon-40 nuclei and the target was carbon-12. The kinetic energy of the beam was 300 AMeV. We have assumed 50,75, and 100ps timing resolution for the MRPCs in the simulation. It was found that the simulation fit the experimental result best with a timing smearing of 50ps. See Fig.\ref {simulation}. The top (a) plot shows the deduced $T_0$ time resolution by a single track, while the bottom (b) plot illustrates its dependence on the number of tracks that hit the T0 detector. The collision vertex was fixed, so its position did  not contribute to the overall resolution. By comparing the $T_0$ time resolution with two tracks in Fig. \ref {simulation} (bottom b) and in Fig.\ref {time6} (bottom b), the MRPC time resolution, including contributions from electronics, particle momentum variation and magnetic field, should be <50ps. This is consistent with the results for IHEP-E3 beam test and validated the excellent performance of the T0 prototype, which completely fulfils the design goal.\par
Fig. \ref {simulation} (bottom b) also shows that the T0 time resolution quickly decreased to $\sim$50 ps for more than five track, and it was saturated at $\sim$30 ps. We would expect even better performance by fine-tuning and calibration of the experiment. For comparison, the T0 time resolution with a one-layer design for the T0 detector is also shown, it is $\sim$30-50\% worse than the double-layer design. This confirmed our expectation when designing this detector.

\section{Conclusions}
Based on MRPC technology, a prototype CSR-T0 detector (of 1/4 acceptance) was designed and produced. The MRPCs, FEE and the timing performance of the T0 prototype have been tested with a hadron beam and a heavy-ion beam. The efficiency of a single MRPC module was also examined, but the overall trigger efficiency for a T0 detector could not be quantified owing to the lack of beam counting rate information. A GEANT4-based simulation was done to evaluate the experimental data analysis results. The intrinsic time resolution of an MRPC, including electronics' contribution, was found to be <50 ps for charged hadrons at a momentum of <1 GeV/c. From the heavy-ion beam test at CSR, the timing performance of the T0 prototype has been evaluated and met our expectation, which suggests that our high expectations for a full-coverage T0 are promising.\par
However, the evaluation of the efficiency of the T0 detector has a real problem, because we do not have record of the beam-target reaction rate, so we have no point of reference. We can only estimate the efficiency from a simulation and the efficiency of a single MRPC. One thing we may do is to find the global time dependence of the T0 trigger and compare it to the beam luminosity to ascertain whether their time structure are the same ($\sim$2-s beam spill per 30 s).

%
%

%
%
%
%
%

\begin{acknowledgements}
We thank the E3 test beam group, IHEP, Beijing and the CSR beam group, Institute of Modern Physics (IMP), Lanzhou for their support to conduct the beam tests successfully. This project is supported by the National Natural Science Foundation of China (U1232206), the International Cooperation and Exchanges Project of NSFC (11420101004) , the National Program on Key Basic Research Project of China-973 Program (2015CB856902), and the CAS Center for Excellence in Particle Physics (CCEPP).
\end{acknowledgements}


\begin{thebibliography}{9}

\bibitem{myers1998nuclear}
W. Myers, W. Światecki, Nuclear equation of state, Physical Review C
57 (6) (1998) 3020.
\bibitem{braun2007quest}
P. Braun-Munzinger, J. Stachel, The quest for the quark–gluon plasma,
Nature 448 (7151) (2007) 302.
\bibitem{zhang2010searching}
S. Zhang, J. Chen, H. Crawford, D. Keane, Y. Ma, Z. Xu, Searching for on-
450set of deconfinement via hypernuclei and baryon-strangeness correlations,
Physics Letters B 684 (4-5) (2010) 224–227.
\bibitem{McLerran:2003yx}
L. McLerran, RHIC physics: The Quark gluon plasma and th
condensate: Four lectures, 2003. arXiv:hep-ph/0311028.
\bibitem{mclerran2009quarkyonic}
L. McLerran, Quarkyonic matter and the revised phase diagram of qcd,
Nuclear Physics A 830 (1-4) (2009) 709c–712c.
\bibitem{andronic2010hadron}
A. Andronic, D. Blaschke, P. Braun-Munzinger, J. Cleymans, K. Fukushi-
ma, L. McLerran, H. Oeschler, R. Pisarski, K. Redlich, C. Sasaki, et al.,
Hadron production in ultra-relativistic nuclear collisions: Quarkyonic mat-
ter and a triple point in the phase diagram of qcd, Nuclear Physics A 837 (1-2) (2010) 65–86.
\bibitem{feng2010probing}
Z.-Q. Feng, G.-M. Jin, Probing high-density behavior of symmetry energy
from pion emission in heavy-ion collisions, Physics Letters B 683 (2-3)
(2010) 140–144.
\bibitem{xie2013symmetry}
W.-J. Xie, J. Su, L. Zhu, F.-S. Zhang, Symmetry energy and pion pro-
duction in the boltzmann–langevin approach, Physics Letters B 718 (4-5)
(2013) 1510–1514.
\bibitem{sumiyoshi1994relativistic}
K. Sumiyoshi, H. Toki, Relativistic equation of state of nuclear matter for
the supernova explosion and the birth of neutron stars, The Astrophysical
Journal 422 (1994) 700–718.
\bibitem{steiner2005isospin}
A. W. Steiner, M. Prakash, J. M. Lattimer, P. J. Ellis, Isospin asymmetry
in nuclei and neutron stars, Physics reports 411 (6) (2005) 325–375.
\bibitem{xiao2014probing}
Z.-G. Xiao, G.-C. Yong, L.-W. Chen, B.-A. Li, M. Zhang, G.-Q. Xiao,
N. Xu, Probing nuclear symmetry energy at high densities using pion, kaon,
eta and photon productions in heavy-ion collisions, The European Physical
Journal A 50 (2) (2014) 37.
\bibitem{xiao2009nuclear}
Z. Xiao, L.-W. Chen, F. Fu, B.-A. Li, G. Jin, H. Xu, G. Yong, M. Zhang,
Nuclear matter at a hirfl-csr energy regime, Journal of Physics G: Nuclear
and Particle Physics 36 (6) (2009) 064040.
\bibitem{xiao2014probing1}
Z.-G. Xiao, G.-C. Yong, L.-W. Chen, B.-A. Li, M. Zhang, G.-Q. Xiao,
N. Xu, Probing nuclear symmetry energy at high densities using pion, kaon,
eta and photon productions in heavy-ion collisions, The European Physical
Journal A 50 (2) (2014) 37.
\bibitem{Lu2016}
L. Lü, H. Yi, Z. Xiao, M. Shao, S. Zhang, G. Xiao, N. Xu, Con-
ceptual design of the hirfl-csr external-target experiment, Science China
485Physics, Mechanics $\&$ Astronomy 60 (1) (2016) 012021. doi:10.1007/
s11433-016-0342-x. URL https://doi.org/10.1007/s11433-016-0342-x.
\bibitem{A:2005}
UrQMD, Urqmd, http://urqmd.org/ (2005).
\bibitem{gapienko2013studying}
V. Gapienko, O. Gavrishchuk, A. Golovin, A. Semak, S. Y. Sychkov, Y. M.
Sviridov, E. Usenko, M. Ukhanov, Studying the counting rate capability of
a glass multigap resistive plate chamber at an increased operating temper-
ature, Instruments and Experimental Techniques 56 (3) (2013) 265–270.
\bibitem{yong2012prototype}
S. Yong-Jie, Y. Shuai, L. Cheng, Q. Sen, X. Lai-Lin, F. Zai-Wei, H. Yue-
Kun, C. Hong-Fang, T. Ze-Bo, S. Ming, et al., A prototype mrpc beam test
for the besiii etof upgrade, Chinese Physics C 36 (5) (2012) 429.
\bibitem{zeballos1996new}
E. C. Zeballos, I. Crotty, D. Hatzifotiadou, J. L. Valverde, S. Neupane,
M. Williams, A. Zichichi, A new type of resistive plate chamber: the multi-
gap rpc, Nuclear Instruments and Methods in Physics Research Section A:
Accelerators, Spectrometers, Detectors and Associated Equipment 374 (1)
(1996) 132–135.
\bibitem{spegel2000recent}
M. Spegel, A. Collaboration, et al., Recent progress on rpcs for the alice tof
system, Nuclear Instruments and Methods in Physics Research Section A:
Accelerators, Spectrometers, Detectors and Associated Equipment 453 (1-
2) (2000) 308–314.
\bibitem{anghinolfi2003nino}
F. Anghinolfi, P. Jarron, F. Krummenacher, E. Usenko, M. Williams, Nino,
an ultra-fast, low-power, front-end amplifier discriminator for the time-of-
flight detector in alice experiment, in: 2003 IEEE Nuclear Science Sympo-
sium. Conference Record (IEEE Cat. No. 03CH37515), Vol. 1, IEEE, 2003,
pp. 375–379.
\bibitem{anghinolfi2004nino}
F. Anghinolfi, P. Jarron, A. Martemiyanov, E. Usenko, H. Wenninger,
M. Williams, A. Zichichi, Nino: an ultra-fast and low-power front-end am-
plifier/discriminator asic designed for the multigap resistive plate chamber,
Nuclear Instruments and Methods in Physics Research Section A: Accelera-
tors, Spectrometers, Detectors and Associated Equipment 533 (1-2) (2004)
183–187.
\bibitem{wang201110}
J. Wang, S. Liu, L. Zhao, X. Hu, Q. An, The 10-ps multitime measurements
averaging tdc implemented in an fpga, IEEE Transactions on Nuclear Sci-
ence 58 (4) (2011).
\bibitem{deng2018readout}
P. Deng, L. Zhao, J. Lu, P. Xia, J. Liu, M. Li, S. Liu, Q. An, Readout
electronics of t0 detector in the external target experiment of csr in hirfl,
IEEE Transactions on Nuclear Science 65 (6) (2018) 1315–1323.
\bibitem{lijiacai2004bepc}
L. Jia-Cai, W. Yuan-Ming, et al., A test beam upgrade based on the bepc-
linac, High Energy Phsics and Nuclear Physics 28 (12) (2004) 1269.
\bibitem{besiii2009construction}
B. Collaboration, et al., The construction of the besiii experiment, Nucle-
ar Instruments and Methods in Physics Research Section A: Accelerators,
Spectrometers, Detectors and Associated Equipment 598 (1) (2009) 7–11.
\bibitem{christiansen2004hptdc}
J. Christiansen, Hptdc high performance time to digital converter (2004).
\bibitem{shao2008upgrade}
M. Shao, X. Dong, Z. Tang, Y. Xu, M. Huang, C. Li, H. Chen, Y. Lu,
Y. Zhang, Upgrade of the calibration procedure for a star time-of-flight
detector with new electronics, Measurement Science and Technology 20 (2)
(2008) 025102.
\bibitem{hu2017t0}
D. Hu, M. Shao, Y. Sun, C. Li, H. Chen, Z. Tang, Y. Zhang, J. Zhou,
H. Zeng, X. Zhao, et al., A t0/trigger detector for the external target
experiment at csr, Journal of Instrumentation 12 (06) (2017) C06010.
\bibitem{yang2014test}
S. Yang, Y. Sun, C. Li, Y. Heng, S. Qian, H. Chen, T. Chen, H. Dai,
H. Fan, S. Liu, et al., Test of high time resolution mrpc with different
readout modes for the besiii upgrade, Nuclear Instruments and Methods
in Physics Research Section A: Accelerators, Spectrometers, Detectors and
Associated Equipment 763 (2014) 190–196.
\bibitem{akindinov2004results}
A. Akindinov, A. Alici, F. Anselmo, P. Antonioli, M. Basile, G. C. Romeo,
L. Cifarelli, F. Cindolo, A. De Caro, S. De Pasquale, et al., Results from
a large sample of mrpc-strip prototypes for the alice tof detector, Nucle-
ar Instruments and Methods in Physics Research Section A: Accelerators,
Spectrometers, Detectors and Associated Equipment 532 (3) (2004) 611–
621.
\bibitem{agostinelli2003geant4}
S. Agostinelli, J. Allison, K. a. Amako, J. Apostolakis, H. Araujo, P. Arce,
M. Asai, D. Axen, S. Banerjee, G. Barrand, et al., Geant4—a simulation
toolkit, Nuclear instruments and methods in physics research section A:
Accelerators, Spectrometers, Detectors and Associated Equipment 506 (3)
(2003) 250–303.
\bibitem{gonzalez2005effect}
D. Gonzalez-Diaz, D. Belver, A. Blanco, R. F. Marques, P. Fonte,
J. Garzón, L. Lopes, A. Mangiarotti, J. Marín, The effect of temperature
on the rate capability of glass timing rpcs, Nuclear Instruments and Meth-
ods in Physics Research Section A: Accelerators, Spectrometers, Detectors
and Associated Equipment 555 (1-2) (2005) 72–79.


%
%
%
%
%
%
%
%
%

\end{thebibliography}
\end{document}